\documentclass[
reprint,
aps,
pra,
superscriptaddress,
 amsmath,amssymb,
floatfix,
]{revtex4-2}

\usepackage[colorlinks=true,citecolor=blue, linkcolor=black]{hyperref}
\usepackage{braket}
\usepackage{placeins}

\newcommand{\Tr}{\operatorname{Tr}}
\newcommand{\ketbra}[2]{\lvert #1 \rangle\!\langle #2 \rvert}
\usepackage{graphicx}
\usepackage{dcolumn}
\usepackage{bm}
\usepackage{xcolor}
\usepackage{amsfonts}
\usepackage{stmaryrd}
\usepackage{mathtools}
\usepackage{amssymb}
\everymath{\displaystyle}
\newcommand{\h}[1]{\hat{#1}}

\begin{document}
\title{A multilevel tensor network compression technique for simulating Lindblad dynamics in superconducting circuits}
\author{Adrien Moulinas}
\email{adrien.moulinas@alice-bob.com}
\affiliation{Alice \& Bob, 53 Bd du Général Martial Valin, 75015 Paris, France}
\affiliation{Univ. Grenoble Alpes, CEA, Grenoble INP, IRIG, Pheliqs, 38000 Grenoble, France}

\author{Xavier Waintal}
\affiliation{Univ. Grenoble Alpes, CEA, Grenoble INP, IRIG, Pheliqs, 38000 Grenoble, France}

\date{\today}
\begin{abstract}
Designing superconducting quantum hardware requires simulation tools that can account for various deviations from ideal scenarios. This, in turn, requires approaches that automatically detect certain structures and leverage them to make the computation affordable. Here, we develop a tensor network based technique to simulate the Lindblad dynamics of a few interacting bosonic modes with a focus on superconducting quantum circuits.
The technique detects and takes advantage of two very common situations: 
(i) the density matrix being  pure or not far from pure and 
(ii) the entanglement between different modes being moderate (typically qubit-like). 
However, (iii) the occupation of the modes can be arbitrarily high (making naïve truncations inefficient). To leverage these features, we use three different nested levels of tensor network compression: (i) we work with a global purification of the density matrix, (ii) we compress the connection between different modes to account for the moderate entanglement and (iii) we use a quantics representation of the Fock occupation number. We showcase the technique for the simulation of large cat qubits as well as for the ionization of transmon qubits, demonstrating orders-of-magnitude speed-up with respect to brute force approaches. In the latter example, it brings the simulation, previously reported on a large supercomputing infrastructure, to laptop level. The favorable scaling with system size should bring genuine computer assisted design of these systems within scope.
\end{abstract}

\maketitle

\section{Introduction}\label{sec:intro}
There is something fascinating in the versatility of superconducting circuits: thanks to the protection of the superconducting gap, these 
assemblies of capacitors and (sometimes non-linear) inductors behave 
precisely as the quantum version of their classical dynamics would imply
\cite{Blais2021}. It follows that, in contrast to say semiconducting devices 
whose behavior depends strongly on details at the atomic scale \cite{Martinez2022}, the modeling of these systems has a high predictive power. One can actually \emph{design} quantum Hamiltonians and superconducting circuits now form a leading platform for building a quantum computer \cite{Blais2021,Clarke2008}
with many different qubit designs \cite{Koch2007,Manucharyan2009,Kiyooka2025}.
A crucial element of the modeling is to take into account the
environment that surrounds the bosonic modes themselves.
Indeed, effects such as leakage to higher energy levels, dissipation, and decoherence play a huge role in the fidelity of quantum operations and must be considered \cite{Devoret2013,Wood2018}. Moreover, a recent class of superconducting circuits known as bosonic codes actually use dissipation as a resource to stabilize robust qubits
\cite{Grimm2020,Mirrahimi2014,Guillaud2023}.
The standard way to account for these effects is to describe the dissipative dynamics at the Lindblad equation level \cite{Lindblad1976, Breuer2002, Lidar2019}:
\begin{equation}
\label{eq:lindblad}
\frac{d \h{\rho}}{d t} = -i [\h{H}, \h{\rho}] + \sum_{k=1} ^l\left( \h{L}_k \h{\rho} \h{L}_k^\dagger - \frac{1}{2} \left\{ \h{L}_k^\dagger \h{L}_k, \h{\rho} \right\} \right)
\end{equation}
where $\h{\rho}$ is the density matrix of the system, $\h{H}$ is its Hamiltonian, and $\h{L}_k$ are the Lindblad operators that describe the dissipative processes (with $\hbar = 1$). Our capacity to engineer efficient new circuits is
directly linked to our capacity to integrate this equation and understand the associated physics. As the devices get more complex, numerical tools 
("digital twins" of the experiments) become increasingly more important for device design and experimental data analysis. Efficient direct solvers such as the QuTiP \cite{Lambert2026} and Dynamiqs \cite{Guilmin2025} packages in Python or QuantumOptics.jl in Julia \cite{Kramer2018} have been developed for this task. Unfortunately the curse of dimensionality severely limits our ability to simulate the Lindblad equation beyond the smallest systems. Indeed, given a system of $M$ bosonic modes, each described with a truncated Hilbert space of $N$ states per boson, the memory footprint to store the density matrix is  $N^{2M}$ which quickly becomes prohibitive as $N$ and/or $M$ increase.

\begin{figure}[ht!]
    \centering
    \includegraphics[width=1\linewidth]{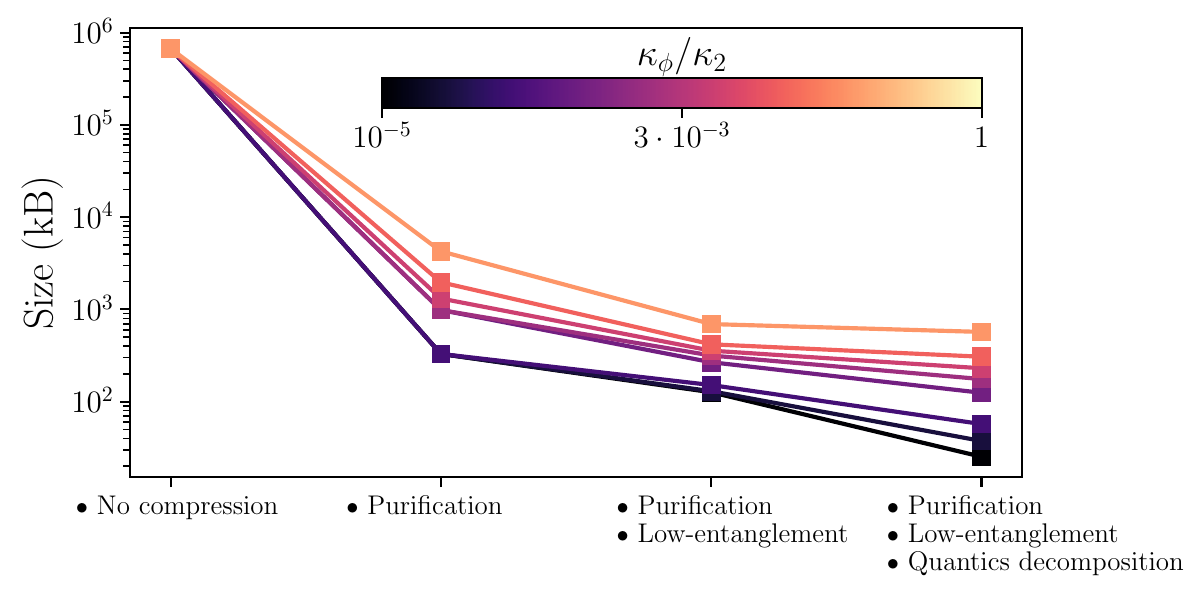}
    \caption{\textbf{Role of the three axes of compression for a cat qubit.} The plot shows the size of the compressed density matrix versus which compression stages are activated. The compression ratio depends significantly on the ratio between the strength of the dephasing $\kappa_\phi$ and the photon loss rate $\kappa_2$ of the buffer. Bad cat qubits (large 
 $\kappa_\phi$) have a more complex dynamics which is more expensive to simulate. See section \ref{sec:compression} for more details.} 
    \label{fig:compression_cat}
\end{figure}

To make progress, one needs some sort of compression. A natural approach is to use tensor network techniques to represent quantum states in a low-rank format using the so-called Matrix Product States (MPS)\cite{Oseledets2011, Schollwock2011, Orus2014}. Important efforts are being devoted to extend these techniques to Lindblad dynamics  \cite{Zwolak2004}. The simplest approach for this is to replace the MPS representing a wave-function with a MPO representing a density matrix \cite{Oh2021, Noh2020, Prosen2009}. This comes with a possible loss of the positive semidefinite property of the density matrix during the truncation as well as a need of global operations to normalize the state or obtain expectation values of operators. Another approach is to use an ansatz more suited to represent density matrices, sometimes referred to as a Matrix Product Density Operator (MPDO) or Locally Purified Tensor Network (LPTN) \cite{Verstraete2004b, Jaschke2018, Werner2016, Muller2024}. This allows both to restrict the ansatz to positive semidefinite matrices and to give a way to control its mixedness. 
In the present work, we build and improve on these approaches using three axes of compression controlled by three parameters called "bond dimensions" $\chi_\mu$, $\chi_e$ and $\chi_q$.
\begin{itemize}
    \item Purity. We use a global purification of the density matrix in terms of $\chi_\mu$ pure states bringing the memory footprint from $N^{2M}$ down to
    $\chi_\mu N^{M}$. This is particularly efficient for e.g. the simulation of cat qubits where the purity must stay high for the device to be in an interesting regime. 
    \item Low-entanglement. This is a standard step in tensor network techniques: one compresses the state according to the amount of entanglement of one bosonic mode with respect to the others. This  brings the memory footprint from $\chi_\mu N^{M}$ down to $\sim \chi_\mu \chi^2_e N M$. 
    When the bosonic modes actually implement qubits, $\chi_e$ increases at most as $\chi_e\sim 2^{M/2}$ (possibly much less if the qubits are moderately entangled) so this step potentially represents a huge compression.
    \item Quantics. This aspect is perhaps the least standard contribution of the present work. We compress the local Hilbert space of a bosonic mode by writing the Fock occupation $n$ (occupation of the (an)harmonic oscillator) in its binary format $n= 2^{R-1} n_1 + 2^{R-2} n_2 +\cdots n_R$ (where the $n_i$ are binary numbers and $N\approx 2^R$) and treat the variables $n_i$ as separate variables \cite{Oseledets2009,Khoromskij2011}. This  brings the memory footprint from $\chi_\mu \chi^2_e N M$ down to $\sim \chi_{\mu} \chi^2_e \chi_q^2 (\log_2 N) M$. This is highly relevant for applications which explore large values of $N$.
\end{itemize}
Taken together, these different levels of compression provide orders of magnitude reduction of the memory footprint and computing time in the regimes for which they have been designed. An example is shown in 
Fig. \ref{fig:compression_cat} for a system that will be discussed in detail below (two bosonic modes forming a cat qubit).

This paper is organized as follows: 
Section \ref{sec:fock_qtt} introduces our quantics 
\cite{Ritter2024, Fernandez2025} approach to represent photon number, together with some basic tensor network definitions. We discuss which representation we think is more favorable.
Section \ref{sec:schrodinger} demonstrates the integration of the Schrödinger equation in this format and studies various integration schemes. 
In Section \ref{sec:compression}, we turn to the Lindblad equation and construct the three-fold compression of the density matrix explicitly. Section \ref{sec:schemes} presents our numerical scheme to integrate the Lindblad equation in this format. Finally, we present the results of our method on two different examples: a Z gate on a large cat state in section \ref{sec:cat}, and the ionization of a transmon qubit coupled to a strongly driven cavity in section \ref{sec:transmon}. We show that our method is able to simulate those systems with a much more favorable scaling with system size than existing methods.

\section{A compact representation of bosonic Hamiltonians in the Fock basis with Tensor Trains}\label{sec:fock_qtt}

In this section, we consider a single bosonic mode described by
a creation (resp. annihilation) operator $\h{a}^\dagger$ (resp. $\h{a}$) with
$[\h{a},\h{a}^\dagger]=1$. The corresponding Hilbert space has an infinite size; the goal of this section is to construct an efficient tensor network representation of it. We have considered two alternative possibilities.
The first is to work in Fock space, i.e. in the eigenbasis $\ket{n}$ of the occupation operator $\h{n} \ket{n} = n \ket{n}$ with $\h{n} = \h{a}^\dagger\h{a}$,
\begin{equation}
    \ket{\psi} \equiv \sum_n \psi_n \ket{n}.
\end{equation}
The second is to work in real space, i.e. in the eigenbasis $\ket{x}$ of the position operator $\h{x} = [\h{a}^\dagger + \h{a}]/\sqrt{2}$,
\begin{equation}
    \ket{\psi} \equiv \int dx \ \psi(x) \ket{x}.
\end{equation}
We have found that, for the studied examples, it is significantly more efficient to work in Fock space, hence this is the choice we will make. However, in this section, we also consider the real space option as well as the conversion from one basis to the other. Finally, we found experimentally that it is also possible to work in a basis more suited to the specific problem and that the high-compressibility of the system still holds; see section \ref{sec:transmon} for such an example.

\subsection{Quantics representation of the wave function}
The Quantics Tensor Train (QTT) representation of the wavefunction $\psi_n$
is deceptively simple: first, one truncates the Hilbert space and keeps
an exponentially large number $2^R$ of Fock states; second, one writes the integer $n$ in its binary form
$n = \sigma_1\sigma_2...\sigma_R$ with $\sigma_i \in  \{0,1\}$ or more explicitly,
\begin{equation}
n(\{\sigma_i\}) = \sum_{i=1}^{R} 2^{R-i} \sigma_i.
\end{equation}
This step transforms the exponentially large vector $\psi_n$ into
a tensor with $R$ indices defined as,
\begin{equation}
\Psi_{\sigma_1,\ldots,\sigma_R} \equiv \psi_{n(\{\sigma_i\})}.
\end{equation}
Third, we factorize this tensor into a Tensor Train (TT) also known as Matrix Product State (MPS) format, using a rank-revealing method to express it as a chain of 3-dimensional tensors. More specifically, such a representation is defined as:
\begin{equation}
    \Psi_{\sigma_1,\ldots,\sigma_R}=\sum_{\{a_i\}} M_{a_1}^{(1)}(\sigma_1) M_{a_1,a_2}^{(2)}(\sigma_2)\cdots M_{a_{R-1}}^{(R)}(\sigma_R)
    \label{eq:MPS}
\end{equation}
where each matrix $M^{(i)}(\sigma)$ is a $\chi_{i-1}\times \chi_{i}$ matrix with all the bond dimensions $\chi_i\le \chi_q$ smaller than the quantics bond dimension $\chi_q$. In a first generation of algorithms, this representation was either constructed analytically or numerically using SVD-based methods \cite{Fannes1992,Bridgeman2017,Oseledets2011, Oseledets2010b,Khoromskij2011,Waintal2026,Frankenbach2025,Holscher2025}.

However, the latter scales exponentially with the size of the system and requires having a full-matrix representation of the system \cite{DeLathauwer2000}. Recently, a new class of algorithms, that we shall refer to collectively as "Tensor Cross Interpolation" (TCI) has been developed to build these TT in polynomial time \cite{Oseledets2010a, Savostyanov2014,Fernandez2025}.

\begin{figure}[ht!]
    \centering
    \includegraphics[width=0.9\columnwidth]{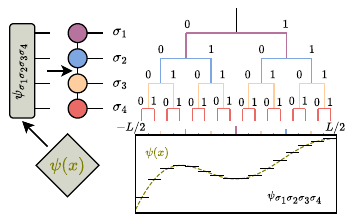}
    \caption{\textbf{Illustration of the QTT framework.} Starting with a function $\psi$ of a continuous variable $x$ defined over an interval $[a,b]$, we discretize it with $2^R$ points ($R=4$ here) and transform it into an R-dimensional tensor $F$. We then use the TCI algorithm to write this tensor as a tensor train with $R$ tensors.}
    \label{fig:Quantics}
\end{figure}

An analogous QTT construction can be done for $\psi(x)$.  We consider an exponentially large box of size $L$ and discretize it in $2^R$ equally spaced points. The quantics discretization reads, 
\begin{equation}
x(\{\sigma_i\}) = -\frac{L}{2} + L\sum_{i=1}^{R} \frac{\sigma_i}{2^i} 
\end{equation}
which allows one to define the tensor 
$ \Psi_{\sigma_1,\ldots,\sigma_R}' = \psi(x(\{\sigma_i\}))$. The rest of the construction is identical to the Fock space one. A schematic of the QTT construction is shown in Fig.\ref{fig:Quantics}.

\subsection{Construction of the Matrix Product Operators}\label{sec:mpo_build}

The next step is to construct efficient representations of the operators that one will have to apply on the different states. Mostly, we deal with operators of the form $\h{a}^m$,  $(\h{a}^{\dagger})^m$ or combination thereof. The Matrix Product Operators (MPOs) are to exponentially large matrices what MPSs are to exponentially large vectors; we refer to \cite{Schollwock2011} for an introduction. Each matrix $M^{(i)}(\sigma)$ that defines a MPS becomes of the form $M^{(i)}(\sigma\sigma')$ for a MPO \cite{Orus2014,Schollwock2011,Waintal2026}.

\subsubsection{\texorpdfstring{MPO of the annihilation operator $\h{a}$}{MPO of the annihilation operator a}}
We want to construct a MPO for the exponentially large matrix $A$ defined
as,
\begin{equation}
A_{nn'} \equiv \bra{n}\h{a}\ket{n'} = \delta_{n+1,n'}\sqrt{n'}
\end{equation}
where  $n$ and $n'$ are written in bit format.
There is no exact QTT representation of the function $\sqrt{n}$. However, we can construct an approximate one which converges exponentially with $\chi_q$.
This is done using TCI with the function 
$f:n \mapsto \sqrt{n}$ for $n \in \{0...2^R-1\}$.
\hypertarget{sentence:diagonal}{
This QTT is then converted to a diagonal MPO $B$. This is a simple process where each tensor $M^{(i)}(\sigma)$ is replaced by 
$B^{(i)}(\sigma,\sigma')= M^{(i)}(\sigma)\delta_{\sigma\sigma'}$.}
Next, we need a MPO for the shift operator $T_{ij} = \delta_{i+1,j}$. This is a standard MPO that can be constructed, for instance, using the magic tensor described in \cite{Waintal2026}. The sought MPO $A$ is equal to 
$A = TB$ and can therefore be obtained with a standard MPO$\cdot$MPO multiplication.  

\begin{figure}[ht]
    \centering
    \includegraphics[width=1\linewidth]{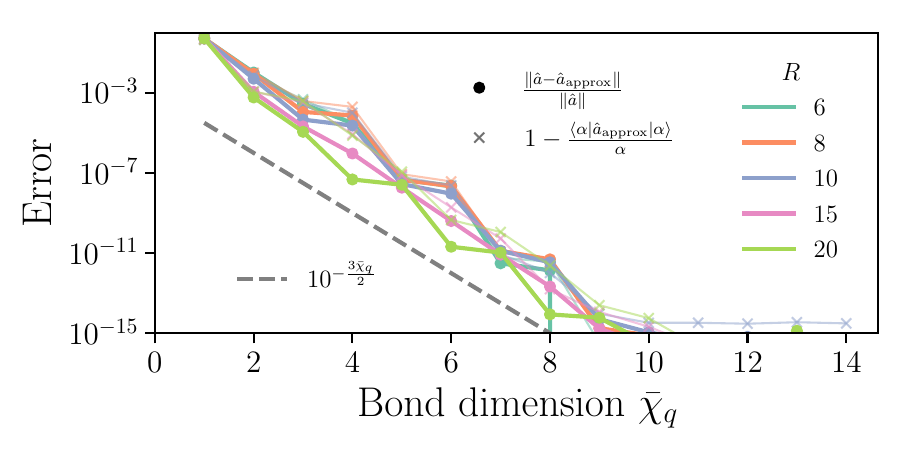}
    \caption{\textbf{Construction of $\h{a}$ in QTT form.} Evolution of the error of the MPO representing the annihilation operator $\h{a}$ with the bond dimension $\overline{\chi}_q$ for various $R$ values. We observe that a bond dimension of $10$ is enough to reduce both errors under $10^{-14}$, for all values of $R$.}
    \label{fig:building_a}
\end{figure}

Fig. \ref{fig:building_a} shows the error we have obtained versus
the MPO rank $\bar\chi_q$ (we use a bar to differentiate the ranks of MPOs
from those of QTT). We observe an extremely fast convergence $\sim 10^{-3\bar\chi_q/2}$ of the error for all values of the number of bits $R$.
The figure shows two metrics for the error. The first is the normalized distance between the built MPO and the exact matrix. The second is the value of $\langle \alpha | \h{a}_{\mathrm{approx}} | \alpha \rangle /{\alpha}$, where $| \alpha \rangle$ is a very large coherent state with $\alpha = 2^{\frac{R-1}{2}}$. This value should ideally be equal to $1$. We find that, for any size of the tensor, a MPO of rank $10$ is enough to represent this operator up to numerical precision.

\subsubsection{Other MPOs}
The construction of other MPOs is done following a similar approach.
The MPO of $\h{a}^\dagger$ is given by $T^\dagger B$ using the above notations. The MPO of powers of the form $\h{a}^m$ can also be built 
with the same technique. We note that 
    \begin{equation}
        \bra{n}\h{a}^m\ket{n'} = \delta_{n+m,n'}\sqrt{\frac{n'!}{n!}}.
    \end{equation}
Hence the MPO requires a generalized shift operator (whose construction is explained in \cite{Waintal2026}) and a TCI MPS for the function 
$f:n \mapsto \sqrt{n!}$ and its inverse (or simply of the falling factorial).
Similarly, we have 
\begin{align}
\bra{n}(\h{a}^\dagger)^m \h{a}^m\ket{n'}  &=& \delta_{n,n'}\prod_{p=0}^{m-1}(n-p) \nonumber \\ 
&=& \delta_{n,n'}\sum_{p=0}^{m}s(m,p)n^p
\end{align} 
where $s(m,p)$ are the Stirling numbers of the first kind. 
It follows that $(\h{a}^{\dagger})^m\h{a}^m$ can be constructed exactly with rank $m+1$. Indeed, polynomials admit exact quantics representations, and we can promote the exact MPS into a MPO using the \hyperlink{sentence:diagonal}{above-mentioned trick}. 
Generalization to operators that act on multiple systems such as 
$\h{a}^m (\h{b}^\dagger)^n$ is obtained straightforwardly by applying the two MPOs of $\h{a}^m$ and $(\h{b}^\dagger)^n$ on the separate parts of the tensor network. Last, linear combinations or products of the above MPOs can be built using the standard toolbox to add and multiply MPOs.

Table \ref{tab:ranks_mpo} summarizes the quantics bond dimension needed to represent different bosonic operators, irrespectively of the Hilbert space size.

\begin{table}[ht]
    \centering
    \begin{tabular}{||c|c|c||}
        \hline   
        Operator & Rank & Precision \\
        \hline \hline
        $\h{a}$& $8$ & $10^{-12}$ \\ \hline
        $\h{a}+ \h{a}^\dagger$& $10$ & $10^{-12}$ \\ \hline
        $\h{a}^n$& $8$ & $10^{-12}$\\ \hline
        $\h{a}^{\dagger m}\h{a}^m$& $m+1$ & Exact\\ \hline
    \end{tabular}
    \caption{\textbf{Ranks of different bosonic operators in QTT form.} All operators admit a QTT representation of low rank, independent of the system size.}
    \label{tab:ranks_mpo}
\end{table}

\subsection{Conversion between the Fock and position representation}
\begin{figure}[ht!]
    \centering
\includegraphics[width=\columnwidth]{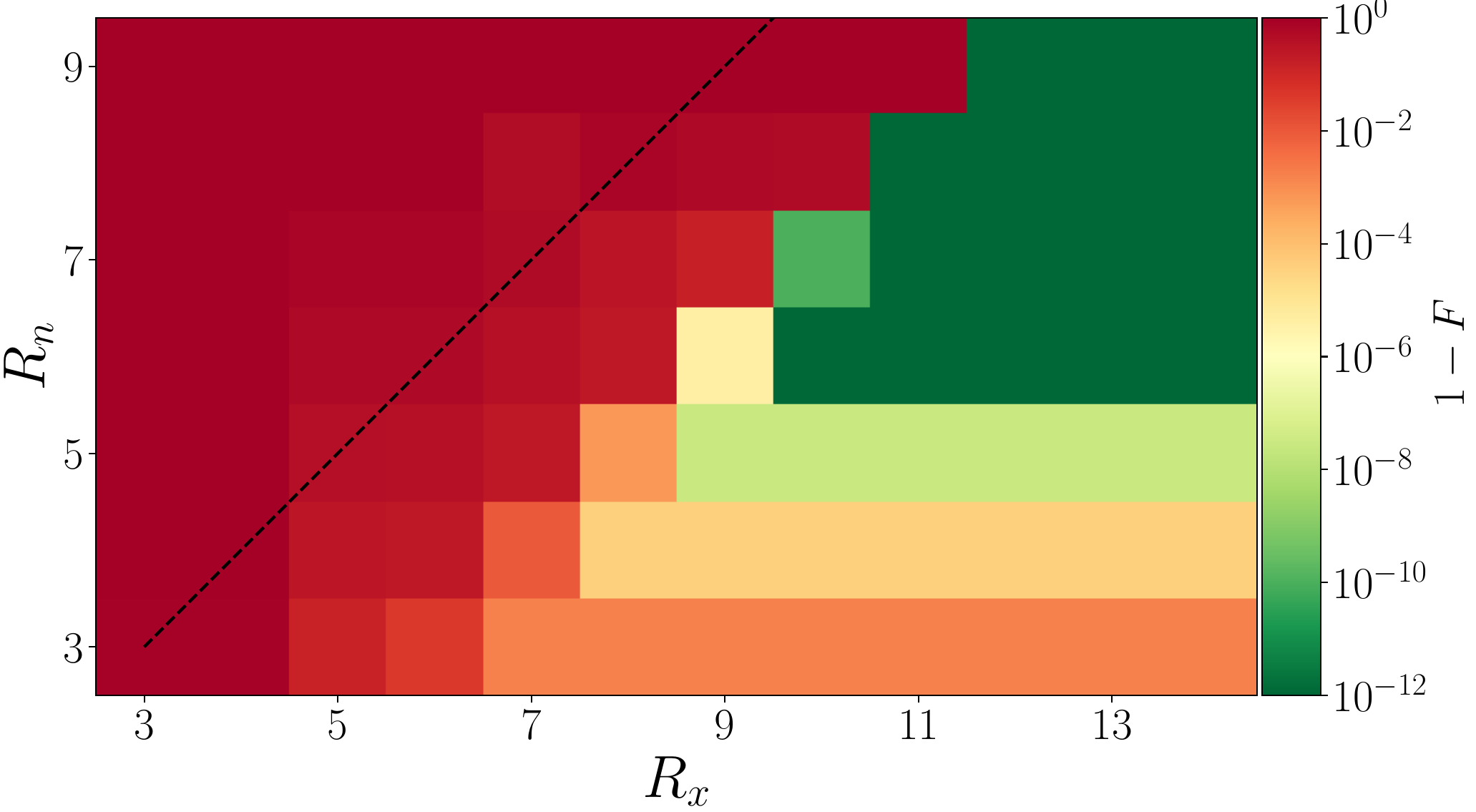}
    \caption{\textbf{Fidelity $F$ of the conversion between Fock and position representation.}  Fidelity of the conversion versus number of bits in position ($R_x$) and Fock ($R_n$) space. A coherent state $|\alpha\rangle$ with $\left| \alpha \right|^2 = 2^{R_n}/3$ is mapped from Fock space to position space and back. The dotted black line corresponds to $R_x = R_n$ and serves as a guide to the eye.
    }
    \label{fig:fid_fock_pos}
\end{figure}

The two representations are simply related by

\begin{align}
\label{eq:basis_change}
\psi(x) =& \sum_n \psi_n \ \phi_{n}(x)  \\
\psi_n  =& \int dx\ \psi(x) \phi_{n}(x)
\end{align}
with
\begin{equation}
\label{eq:psi_n}
\phi_{n}(x)=\frac{\pi^{-\frac{1}{4}}}{\sqrt{2^{n}n!}}
H_{n}\left(x\right) e^{-x^2/2}
\end{equation}
where $H_n(x)$ is the Hermite polynomial of order $n$. It turns out that
Eq.\eqref{eq:psi_n}, seen as a large matrix with $x$ as rows and $n$ as columns, admits a low rank MPO representation that we have calculated using TCI. Below we discuss two numerical experiments whose results have 
motivated us to choose to work in the Fock basis in the rest of this article.

In the first, we have constructed the MPS of a coherent state 
$|\alpha\rangle$ in Fock space, then used Eq.\eqref{eq:basis_change}
to obtain the position space representation and then back to Fock space. We have measured the fidelity of the overall process versus the number of bits $R_n$ and $R_x$ used respectively in the Fock and position representations.
The results are shown in Fig.\ref{fig:fid_fock_pos}. We conclude that to represent the same state, a smaller number of bits are needed in Fock space compared to position space (green region of the diagram). 

\begin{figure}[ht!]
    \centering
    \includegraphics[width=0.9\columnwidth]{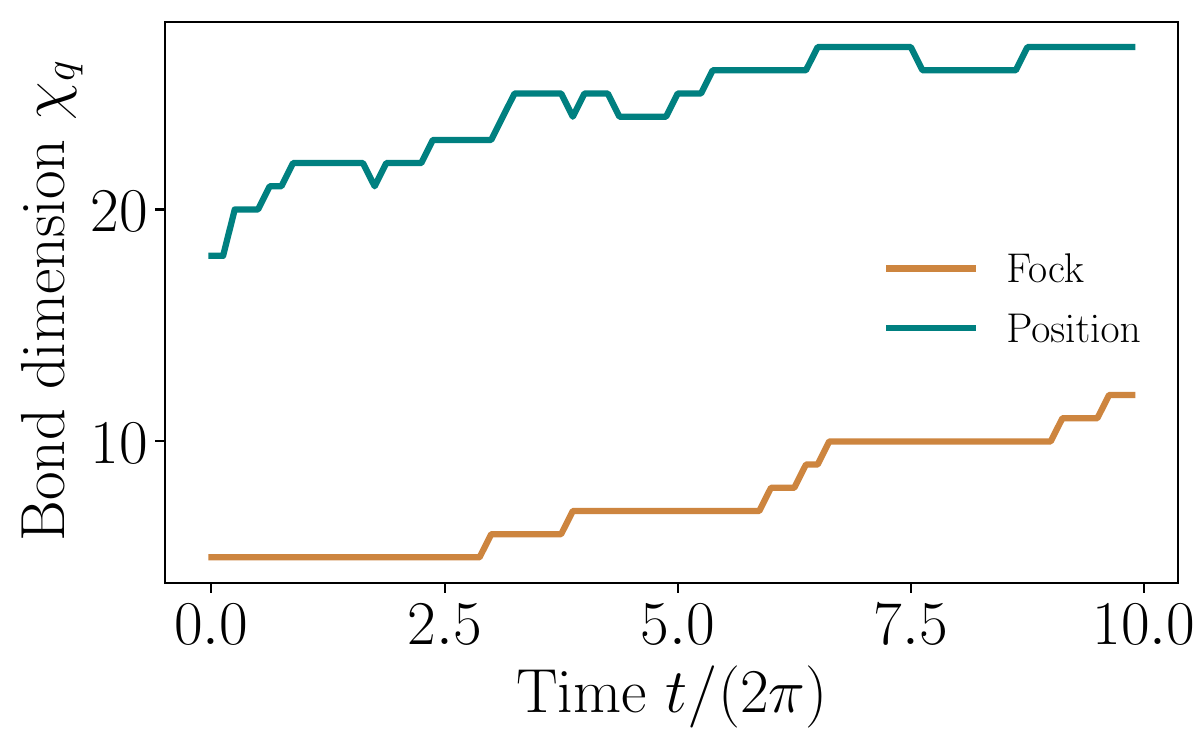}
    \caption{\textbf{Comparison of QTT bond dimension between the Fock and position basis.} The time evolution is simulated using the Magnus-TDVP algorithm described in \ref{sec:magnus_tdvp} for the problem described in \ref{sec:kerr_cavity}. We consider QTT of sizes $R_n = 10$ and $R_x = 12$. We observe that the quantics bond dimension $\chi_q$ is significantly smaller in the Fock space.}
    \label{fig:fock_pos}
\end{figure}

In the second experiment, we have performed a Hamiltonian time-evolution simulation (as explained in the next section). Fig. \ref{fig:fock_pos} presents the evolution of the bond dimension versus time for a driven anharmonic oscillator with two-photon drive and Kerr non-linearity 
(see section \ref{sec:kerr_cavity} for the precise description).
 We use a tensor train of size $R_n=10$ for Fock space simulations and $R_x = 12$ for position space simulations. We observe that at all steps of the evolution, the bond dimension of the QTT is higher in the position basis
 by almost a factor two. Also notice that the bond dimension varies very weakly with time in sharp contrast to what is observed in the dynamics of many-body simulations \cite{Gobert2005,Schollwock2011,Jaschke2018,Daley2012}. Indeed, here the bond dimension $\chi_q$ characterizes the interaction between different scales (the $\sigma_i$ variables) which is
\emph{not} the quantum entanglement between different qubits.

\section{Hamiltonian simulations} \label{sec:schrodinger}

We are now in a position to write a given Hamiltonian $\h{H}(t)$ as a MPO
$\mathbf{H}(t)$ in Fock space. We start by studying the Hamiltonian dynamics,
\begin{equation}
i \dfrac{d}{dt}\ket{\psi(t)} = \h{H}(t)\ket{\psi(t)}
\end{equation}
which, written in terms of our QTT objects, takes the form,
\begin{equation}
\label{eq:mpo-mps-dt}
i \dfrac{d}{dt}\Psi(t) = \mathbf{H}(t)\Psi(t)
\end{equation}
The goal of this section is to explore various integration schemes that can be used to obtain the corresponding QTT evolution. Our numerical experiment indicates that the Time-Dependent Variational Principle (TDVP) method \cite{Haegeman2016}, properly extended to account for the fact that $\h{H}(t)$ is time-dependent (Magnus expansion), is superior to more naïve approaches that get in and out of the QTT manifold. Note that the benchmark below was performed chronologically \emph{after} the design of our Lindblad solver so that we have not yet adapted the Magnus-TDVP approach to the Lindblad solver.

\subsection{A survey of different integration schemes.}
Formally, the evolution of the state during a time step of size $h_t$ is given by the evolution operator,
\begin{equation}
  \mathbf{U}(t,h_t)\equiv \mathcal{T}e^{-i \int_t^{t+h_t} \mathbf{H}(t')dt'}
\end{equation}
so that $\Psi(t+h_t) = \mathbf{U}(t,h_t) \Psi(t)$; $\mathcal{T}$ is the usual time-ordering operator. In this section, we compare different integration schemes to
construct a controlled approximation of $\mathbf{U} \Psi(t)$,
\begin{itemize}
\item A global explicit method ($4^\text{th}$-order Runge-Kutta)
\item A global implicit method (Crank-Nicolson)
\item A naïve TDVP method.
\item A modified TDVP method with Magnus expansion.
\end{itemize} 
We also considered a Trotterized evolution that gave results consistent with the first two methods (not shown here, see \cite{Niedermeier2026} for an example).
The first two global methods are different in nature from the TDVP methods: evolution of $\Psi(t)$ as a whole versus evolution of its constituents $M^{(i)}(\sigma)$ one after the other. In the global methods, the evolution mechanically leads to an increase of the QTT rank and therefore requires a recompression of the QTT after each time step. 

\subsubsection{Explicit global Runge-Kutta scheme}
Global methods amount to taking Eq.\eqref{eq:mpo-mps-dt} as if $\Psi(t)$ was a large vector, $\mathbf{H}(t)$ a large matrix and using standard ordinary differential equation solvers to integrate them. The only modification is that we replace the ordinary linear algebra operations (e.g. a matrix vector product) by their tensor network counterparts, 
(e.g. a MPO$\cdot$MPS product). A wide variety of algorithms exist for MPO$\cdot$MPS product \cite{Oseledets2011,AlDaas2023,Schollwock2011,Paeckel2019,Camano2026}. Here we use a combination of the zip-up \cite{Stoudenmire2010} and fitting \cite{Verstraete2004a} algorithm. See \cite{Waintal2026} for the sum of two MPS or solving MPO$\cdot$MPS = MPS linear problems.

The simplest global solver is to use an explicit Runge-Kutta integration \cite{Butcher1996, Runge1895}. Here we have found that a good compromise was
the standard fourth order Runge-Kutta (RK4) scheme. Denoting,
\begin{align}
\Psi_0 &=& \Psi(t) \\
\Psi_1 &=& -i\mathbf{H}(t)\Psi_0\\
\Psi_2 &=& -i\mathbf{H}(t+\tfrac{h_t}{2}) \left( \Psi_0 + \tfrac{h_t}{2} \Psi_1 \right)\\
\Psi_3 &=& -i\mathbf{H}(t+\tfrac{h_t}{2}) \left( \Psi_0 + \tfrac{h_t}{2} \Psi_2 \right)\\
\Psi_4 &=& -i\mathbf{H}(t+h_t) \left( \Psi_0 + h_t \Psi_3 \right),
\end{align}
we just create these vectors using MPO$\cdot$MPS product. The RK4 scheme approximates 
$\mathbf{U} \Psi(t)$ with the expression,
\begin{equation}
\Psi (t+h_t) \approx \Psi_0 + \tfrac{h_t}{6} \left( \Psi_1 + 2\Psi_2 +
2\Psi_3 + \Psi_4 \right)
\end{equation}
which is valid up to fifth order $O(h_t^5)$.
The QTT is recompressed after each time step.
Irrespective of the fact that we are using QTT, explicit solvers are not numerically stable for stiff equations, which in practice implies that very small time steps 
must be used. The largest energy scale in the problem determines the time step $h_t$. For instance, for a simple harmonic oscillator $\h{H} = \omega_0\h{a}^\dagger a$ where we consider a system with up to $N$ photons, the largest energy scale is $N\omega_0$ which implies that we must use $h_t \ll 1/(N\omega_0)$. In the presence of a non-linear Kerr term of the form $\h{H} = \omega_0(\h{a}^\dagger)^2 a^2$, the condition becomes even more stringent, $h_t \ll 1/(N^2\omega_0)$ \cite{Robin2025}.

\subsubsection{Implicit global Crank-Nicolson scheme}
To address this shortcoming, one could use an implicit method such as
Crank-Nicolson \cite{Crank1947}.
\begin{equation}
  \mathbf{U}\left(t+h_t,-\tfrac{h_t}{2}\right) \Psi(t+h_t) = 
  \mathbf{U}\left(t,\tfrac{h_t}{2}\right)\Psi(t)
\end{equation}
This equation is solved at each step using an alternating least-squares solver such as the one provided in \cite{Gelss2022}. Since a very good initial condition is known
($\Psi(t+h_t)$ is close to $\Psi(t)$) these iterative solvers converge in very few iterations. We approximate the evolution operator $\mathbf{U}(t,h_t)$ with its second order Taylor expansion which, we have found, is a good compromise,
\begin{equation}
    \mathbf{U}(t,h_t) \approx 1 -ih_t\mathbf{H}(t) - \tfrac{h_t^2}{2}\mathbf{H}^2(t).
\end{equation}
The method is numerically stable and preserves the norm of the wavefunction. 

\subsubsection{Time-Dependent Variational Principle with Magnus expansion}\label{sec:magnus_tdvp}

An approach that has been shown to be very powerful in the context of many-body dynamics is to find effective equations for $\frac{d}{dt} M^{(i)}(\sigma)$ using the time-dependent variational principle \cite{Haegeman2011, Haegeman2016}, directly working in the tangent space of the QTT manifold. The approach is known as TDVP in the field, and we benchmark it here for quantics. We refer to the original literature for the full set of equations. In a nutshell, one derives an effective $1$-site Hamiltonian so that for each tensor, one must integrate,
\begin{equation}
\label{eq:tdvp}
    i\frac{d}{dt}M^{(i)}(\sigma) = \sum_{\sigma'}\h{H}^\text{eff}_{\sigma\sigma'}M^{(i)}(\sigma')
\end{equation} 
After each time step, one performs a singular value decomposition
\begin{equation}
  [M^{(i)}(\sigma)](t+h_t) = [U^{(i)}(\sigma)](t+h_t) \Lambda(t+h_t) V^\dagger  
\end{equation}
then one must evolve $\Lambda(t+h_t)$ \emph{backward} in time (using another, $0$-site, effective Hamiltonian); one absorbs $\Lambda(t) V^\dagger$ in $[M^{(i+1)}(\sigma)](t)$ and proceeds with the evolution of $M^{(i+1)}(\sigma)(t)$. A key strength of TDVP is its explicit gauge fixing of the MPS tangent space, which turns the projected dynamics into well-defined local evolution equations for the tensors.

Eq.\eqref{eq:tdvp} is an ordinary differential equation that may also be integrated with various integration schemes, e.g. explicit or implicit. In most existing implementations, such as the Pytenet package \cite{Mendl2018} that we have used, the
Hamiltonian is supposed to be time-independent. The corresponding "plain" TDVP result 
induces an additional error of order $\mathcal{O}(h_t^2)$\cite{Blanes2009}.

To avoid this unnecessary extra error, we use an averaged Hamiltonian $H_\Omega$  obtained using a fourth-order Magnus expansion.  One defines the operator $H_\Omega$ such that $e^{-i\mathbf{H}_\Omega h_t} = \mathbf{U}(t,h_t) + \mathcal{O}(h_t^5)$ \cite{McCulloch2025, Blanes2009}. Its expression reads,
\begin{equation}
\begin{split}
  \mathbf{H}_\Omega = &\tfrac{1}{2} \left(\mathbf{H}(t_1) + \mathbf{H}(t_2)\right) \\&+ i\tfrac{\sqrt{3} h_t}{12} \left[\mathbf{H}(t_1),\mathbf{H}(t_2)\right]
\end{split}
\end{equation}
with,
\begin{equation}
  \begin{aligned}
    t_1 &= t + \tfrac{3-\sqrt{3}}{6} h_t \\
    t_2 &= t + \tfrac{3+\sqrt{3}}{6} h_t
  \end{aligned}
\end{equation}
This approach is particularly suitable for our benchmark where we have considered 
a Hamiltonian of the form
\begin{equation}
  \mathbf{H}(t) = \mathbf{H}_0 + f(t) \mathbf{H}_1
\end{equation}
Indeed, for such a problem, it is enough to calculate the sum
\begin{equation}
    \frac{1}{2}\left(\mathbf{H}(t_1) + \mathbf{H}(t_2)\right) = \mathbf{H}_0 + \frac{f(t_1) + f(t_2)}{2}\mathbf{H}_1
\end{equation}
and the commutator 
\begin{equation}
  \left[\mathbf{H}(t_1),\mathbf{H}(t_2)\right] = [f(t_2) - f(t_1)]\left[\mathbf{H}_0,\mathbf{H}_1\right]  
\end{equation}
once at the beginning of the calculation to obtain the Magnus expansion at each step using a simple linear combination (computationally much cheaper than a MPO$\cdot$MPO contraction).

\subsection{Numerical benchmark with a driven oscillator with Kerr non-linearity}\label{sec:kerr_cavity}
For our benchmark, we consider the Hamiltonian of a cavity with a time-dependent two-photon drive and a Kerr non-linearity.

\begin{equation}
\hat{H}(t)
  = \omega_0 \h{a}^\dagger \h{a}
   +\frac{K}{2} \h{a}^{\dagger 2} \h{a}^2
   +  f(t)\left(\h{a}^2 +\h{a}^{\dagger 2} \right) \\
\label{eq:Hamiltonian}
\end{equation}
with a coherent state as initial state $\ket{\psi(t=0)} = \ket{\alpha_0}$. This Hamiltonian is a simple example of a system which is not exactly solvable. In its semi-classical limit, the state remains a coherent state at all times, with $\alpha(t)$ solution of 
\begin{equation}
    i\tfrac{d}{dt}\alpha  = \omega_0\alpha + 2 f(t)\alpha^*  + K|\alpha|^2\alpha.
\end{equation}
This semi-classical approximation is exact in the limit $K=0$ while in the absence of drive ($f(t) = 0$) the Hamiltonian is diagonal in the Fock basis, and can therefore be integrated exactly.  

\begin{figure*}[ht]
    \centering
    \includegraphics[width=\textwidth]{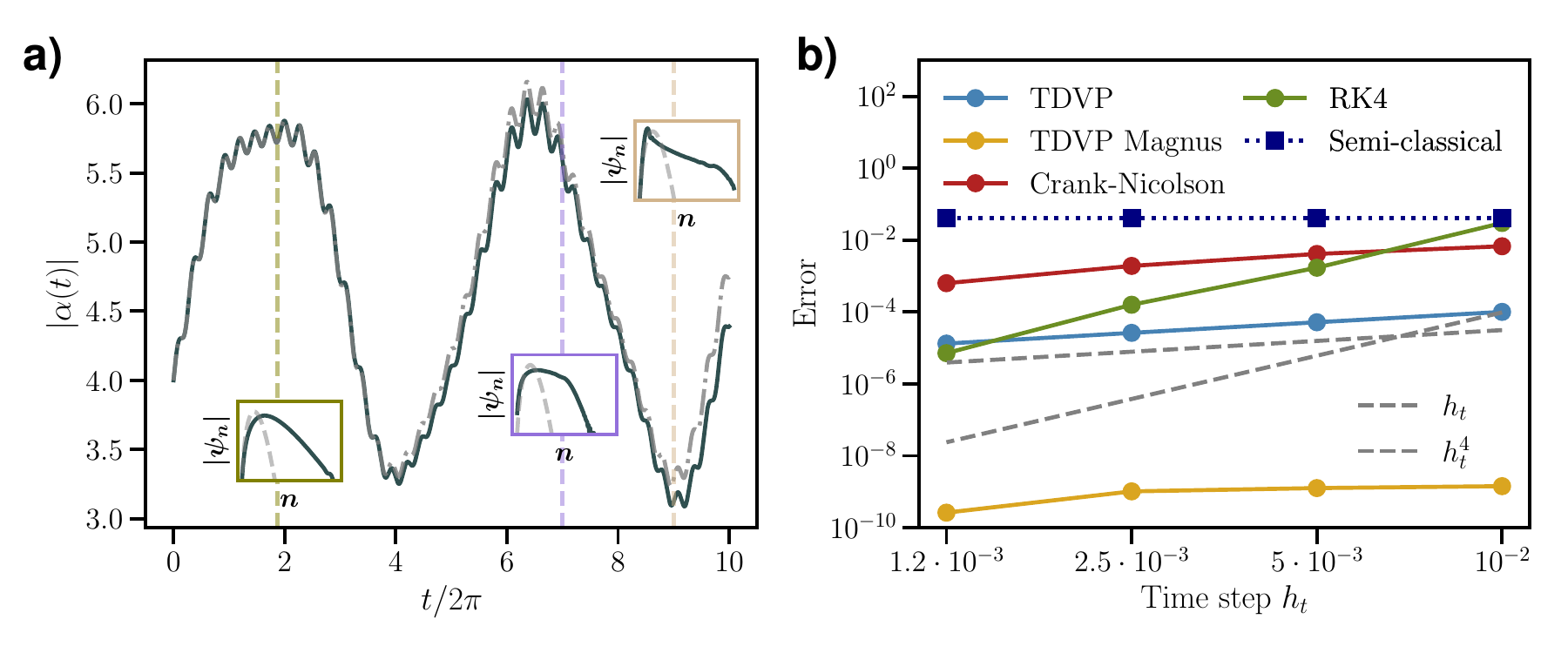}
    \caption{\textbf{QTT simulation of the driven Kerr oscillator described by Eq.\eqref{eq:Hamiltonian}.} See text for the parameters.
    \textbf{a)} Evolution of $\alpha(t) \equiv \bra{\psi (t)} \h{a} \ket{\psi (t)}$ using the exact (black line) and the semi-classical (dash-dot gray line) dynamics.}
The insets represent $\psi_n$ versus $n$ at three different times indicated by the colored vertical dashed lines.
    \textbf{b)} Error $\text{E} = 1 - |\langle \psi_\text{th}(t_f) | \psi(t_f) \rangle|$ versus time step $h_t$ for different methods. The dashed gray lines serve as guides to the eye. The reference calculation of $\psi_\text{th}(t_f)$ was obtained using a plain full-vector solver.
    \label{fig:2-photon drive}
\end{figure*}

In the simulations, we use the following parameters: $R = 8$, $\omega_0 = 1$, $K = \frac{1}{25}$, $\alpha_0 = 4$ with a drive that contains two incommensurable frequencies and amplitudes.
\begin{equation}
    f(t) = \frac{2}{100}\cos{(\sqrt{2} t)} 
         + \frac{2\pi}{100}\cos{(\sqrt{3} t)}.
\end{equation}
All TN operations are performed with a tolerance of $10^{-8}$. 
Fig.~\ref{fig:2-photon drive}a shows the expected value
\begin{equation}
\alpha(t) \equiv \bra{\psi (t)} \h{a} \ket{\psi (t)}
\end{equation}
calculated by integrating the Schr\"odinger equation (black curve) or using
the semi-classical method (gray dash-dot curve). Although the value of $\alpha(t)$ itself obtained in the semi-classical limit is relatively accurate at small time, inspection of the wavefunction itself at different times (insets) reveals that the wavefunction $\psi_n$ differs significantly from a coherent state, even early in the simulation. Fig.~\ref{fig:2-photon drive}b compares the different methods described earlier. As already stated, the TDVP with Magnus expansion appears to be consistently better than other competing approaches. As expected, the variation in time of the Hamiltonian must be explicitly taken into account or an $O(h_t)$ error builds up
($h_t^2$ per time step). The RK4 error indeed scales as $h_t^4$ and is more favorable
in practice than Crank-Nicolson despite being explicit.

\section{Efficient representation of the density matrix} \label{sec:compression}

 In this section, we consider the density matrix $\h{\rho}$ of an open system consisting of $M$ bosonic modes, each with a local Hilbert space of dimension $N$. The density matrix thus has a total of $N^{2M}$ elements. We will show that we can reduce this number significantly by exploiting three compression axes.

The need for such an approach is rather acute, even for relatively small systems.
If we consider the example of cat qubits, the system built by Alice \& Bob, a minimum model uses two modes per qubit. The cavity mode would contain
around $10$ photons which requires a truncation around $N=40$ for accuracy; the buffer mode is mostly empty which requires a smaller truncation, say $N=8$. Overall, a plain description of the corresponding density matrix for two such qubits requires $\sim 16 (40 \times 8)^4 = 160$GB of memory and the associated run time is already prohibitive.
As we shall see this number can be reduced drastically.

\begin{figure*}[ht!]
  \centering
  \includegraphics[width=\textwidth]{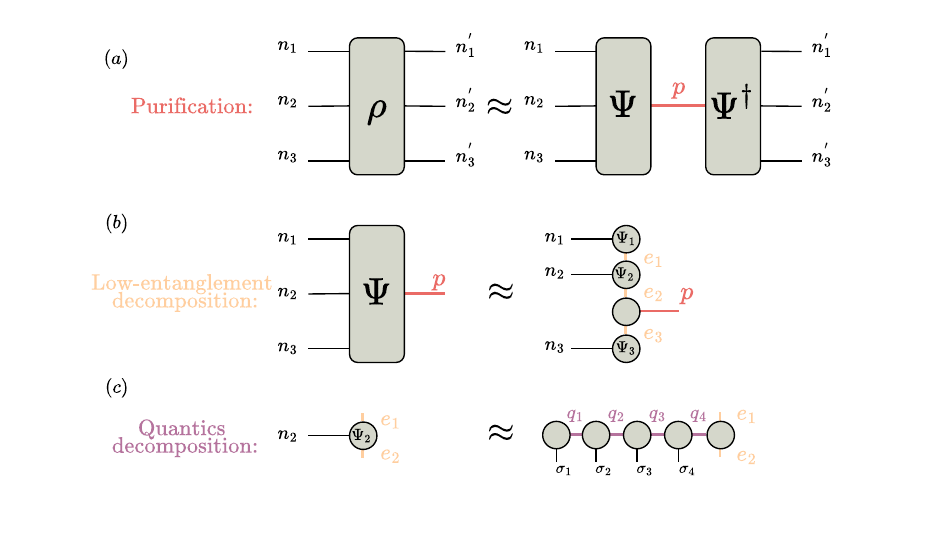}
  \caption{\textbf{Diagrammatic representation of the three compression axes.
  (a)} We start with a density matrix of 3 bosonic systems, each of size $N$, that we approximate as an incoherent sum of $\chi_\mu$ pure states. 
  \textbf{(b)} We store only $\boldsymbol{\Psi}$ in memory, and we assume low entanglement between the different modes, so that the information exchanged between them is quantified by $\chi_e$. 
  \textbf{(c)} We use the quantics compression to further compress each local Hilbert space. We write $N = 2^R$ and use $R$ tensors of bond dimension $\chi_q$ to represent each mode.
  }
  \label{fig:Compression}
\end{figure*}

\subsection{Compression using the expected purity of experimentally relevant states}
We start with a general density matrix $\h{\rho}$ which in Fock space takes the following form,
\begin{equation}
\h{\rho} = \sum_{\mathbf{n}\mathbf{n}'} \rho_{\mathbf{n}\mathbf{n}'} \ket{\mathbf{n}}\bra{\mathbf{n}'}
\end{equation}
where $\mathbf{n}=(n_1,n_2,\dots,n_M)$ is a vector of Fock space occupation. The matrix $\rho_{\mathbf{n}\mathbf{n}'}$ is positive semidefinite and can always be purified, i.e. written as an incoherent sum of pure states,
\begin{equation}
\rho_{\mathbf{n}\mathbf{n}'} = \sum_{\mu=1}^{\chi_\mu(t)} \boldsymbol{\Psi}_{\mathbf{n},\mu} \boldsymbol{\Psi}^\dagger_{\mathbf{n}',\mu}
\end{equation}
where different columns of $\boldsymbol{\Psi}$ are orthogonal. Fig. \ref{fig:Compression}a shows a schematic of this representation which we write in compact form as $\rho = \boldsymbol{\Psi}\boldsymbol{\Psi}^\dagger$ with $\text{Tr} \left( \boldsymbol{\Psi}^\dagger \boldsymbol{\Psi} \right)=1$.

The first axis of compression that we can exploit is that, in many situations of interest, the states that are manipulated remain close to pure states throughout the evolution. This can be quantified using the purity $\text{Tr}[\h{\rho}^2]$ which is equal to 1 for pure states and remains close to 1 for weakly mixed states. Indeed, when the purity drops significantly, it is already the signature that the corresponding state is no longer a resource for quantum computing. This is relevant for e.g. gate simulations (low purity means that decoherence has already polluted the state), bosonic qubit stabilization (dissipation is large but it stabilizes a small manifold of pure states) or readout processes (mixing occurs but only within a few components).

A purity close to unity means in turn that the number of states $\chi_\mu(t)$ kept in the purification can be dynamically truncated to reduce the memory footprint of the density matrix with typically $\chi_\mu(t) \sim 1/\Tr[\rho(t)^2]$.
This compression step reduces the problem size from $N^{2M}$ to $N^M\chi_\mu(t)$ providing an important gain when $\chi_\mu(t) \ll N^M$. Methods using purification
based compression are known as ensemble rank truncation methods \cite{McCaul2021,Appelo2025}. Note that the method is always under control: if the purity is very low, we just need to keep more states until the compression does not give a computational advantage anymore. A secondary advantage of the method is that it guarantees the positivity of the density matrix at all times (in contrast to approaches where one uses e.g. a matrix product operator to represent it).

\subsection{Entanglement-based compression}

The second axis of compression relies on the fact that, in many physically relevant situations, the entanglement between different modes remains low or moderate. 
For instance, in cat qubits, even though the local Hilbert space may require a large
value of $N$, the systems effectively behave as qubits with two states and the entanglement can only grow accordingly. Also, the buffer cavity of these cat qubits
is engineered to remain close to vacuum and therefore carries very limited information about the logical state of the qubit itself \cite{Gautier2023,Putterman2025}.
In other types of qubits such as transmons, one typically engineers the couplings to avoid leakage outside the computational Hilbert space so that, again, the entanglement entropy can grow, at most, as fast $M\log 2$, not $M\log N$.
A similar phenomenon happens in dispersive readout of e.g. a transmon where the cavity acquires information about the qubit state without becoming strongly entangled with it.
Furthermore, dissipation and decoherence mechanisms often act in a way that suppresses the growth of entanglement, either by continuously projecting the system toward a low-dimensional manifold (as in stabilized bosonic codes) or by rapidly removing correlations carried by auxiliary modes.

The standard procedure to take advantage of the limited entanglement between modes is to write the wavefunction as a MPS \cite{Schollwock2011,Eisert2010,Hastings2007,Waintal2026}. We arrive at the format shown in Fig. \ref{fig:Compression}b or explicitly,
\begin{equation}
\boldsymbol{\Psi}_{\mathbf{n},\mu} = \Psi^{(1)}(n_1)\Psi^{(2)}(n_2)\dots
\Psi^{(M)}(n_M) \Psi^{(0)}(\mu)
\end{equation}
where for each $\alpha$, the $N$ matrices $\Psi^{(\alpha)}(n)$ for $\forall n\in \{1\dots N\}$ are at most of size $\chi_e \times \chi_e$. The matrix $\Psi^{(0)}(\mu)$ has a special role; it carries the purification index $\mu$. It may actually be anywhere in the product, not necessarily at the end. This location can be moved around the MPS by contracting the corresponding tensor with its neighbor and then performing a singular value decomposition to factorize it in the chosen direction.

As for the purification formulation, the MPS formulation can always be used. However, it is only advantageous when the entanglement is limited. The effective size of the problem is reduced from $N^M\chi_\mu$ to $(\chi_\mu +NM) \chi_e^2$, which is exponentially more compact provided $\chi_e \ll N^{M/2}$. 

\subsection{Quantics-based compression}
The last compression step consists in using the QTT representation discussed earlier to write each local tensor $\Psi^{(\alpha)}(n)$ with a Hilbert space of size $N$ as a QTT with $R = \log_2(N)$ tensors of bond dimension $\le \chi_q$. We write 
\begin{equation}
\Psi^{(\alpha)}(n)_{e_\alpha e_{\alpha+1}} = M^{(\alpha,0)}_{e_\alpha e_{\alpha+1}}M^{(\alpha,1)}(\sigma_1)\dots M^{(\alpha,R)}(\sigma_R)
\end{equation}
where the "backbone" tensor $M^{(\alpha,0)}$ connects the different modes together,
see the schematics in Fig. \ref{fig:Compression}c. Note that we have the choice between two different orderings of the indices $\sigma_i$: we can place the large scale index $\sigma_1$ close to the backbone MPS or decide to place $\sigma_R$ there instead.
We have observed that the above choice ($\sigma_1$ close to the backbone) yields slightly better compression. Note that here we have focused on the Fock basis, but alternative representations may be better adapted depending on the problem.
Indeed, similar QTT construction could be built in e.g. position space or displaced Fock states \cite{Chamberland2022}. 

\begin{figure}[ht]
    \centering
    \includegraphics[width=1\linewidth]{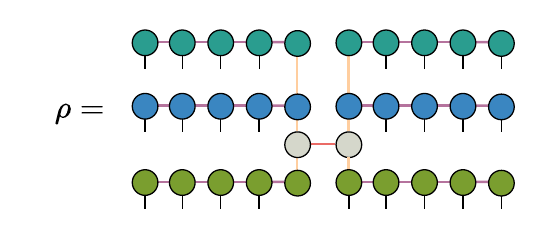}
    \caption{Schematic of the full representation of the system with the three axes of compression. Only the left part of the diagram is stored in memory.}
    \label{fig:full_dm}
\end{figure}

Putting everything together, we arrive at a representation in terms of two coupled "comb" tensor networks as shown in Fig. \ref{fig:full_dm} (remembering that only one comb is kept in memory). 

We can now go back to Fig. \ref{fig:compression_cat} which illustrates the three axes of compression used in our approach. In this figure, we have performed the simulation of a cat qubit undergoing a \hyperref[sec:cat]{Z gate}, for various values of the dephasing rate $\kappa_\varphi$, in the presence of a Lindblad operator $L_\varphi = \sqrt{\kappa_\varphi}\hat{a}^\dagger\hat{a}$. As the dephasing increases, the state becomes more mixed and develops more complex correlations, leading to an increase in both $\chi_\mu$ and the effective bond dimensions $\chi_e$ and $\chi_q$. This results in a larger memory cost, reflecting a reduced compressibility of the state.

\section{QTT Integration of the Lindblad Equation}\label{sec:schemes}

We have the ansatz; we now need to build the surrounding algorithm to integrate 
the Lindblad equation within this representation. At $t=0$, we typically start with
pure and separable states such that $\chi_\mu(0) = \chi_e(0) = 1$ so we need only to construct the QTT representation of each bosonic mode. This is typically done using TCI. Once $\rho$ is put in tensor network format, we perform all calculations within this representation and never leave it.

\subsection{Summing vectors versus summing matrices}

The algorithm that we will build mostly uses standard MPO/MPS technology.
However, there is one subtle point originating from the fact that we're using a
purification of the density matrix, i.e. we work with $\boldsymbol{\Psi}$ not
$\rho = \boldsymbol{\Psi}\boldsymbol{\Psi}^\dagger$.
Consider two such density matrices $\rho_1 = \boldsymbol{\Psi}_1\boldsymbol{\Psi}_1^\dagger$ and $\rho_2 = \boldsymbol{\Psi}_2\boldsymbol{\Psi}_2^\dagger $. We will need to perform two different sorts of tensor network additions: vector additions and matrix additions. In vector addition, we want to construct
\begin{equation}
\boldsymbol{\Psi}_1 + \boldsymbol{\Psi}_2
\end{equation}
while in matrix addition we want to construct 
\begin{equation}
  \rho_1 + \rho_2
    = \boldsymbol{\Psi_1\Psi_1 ^\dagger} + \boldsymbol{\Psi_2\Psi_2 ^\dagger} 
\end{equation} 
which is a very different object since 
\begin{equation}
    (\boldsymbol{\Psi_1 +\Psi_2})(\boldsymbol{\Psi_1 +\Psi_2})^\dagger 
 \neq \boldsymbol{\Psi_1\Psi_1 ^\dagger} + \boldsymbol{\Psi_2\Psi_2 ^\dagger}
\end{equation}

\begin{figure}
    \centering
    \includegraphics[width=1\linewidth]{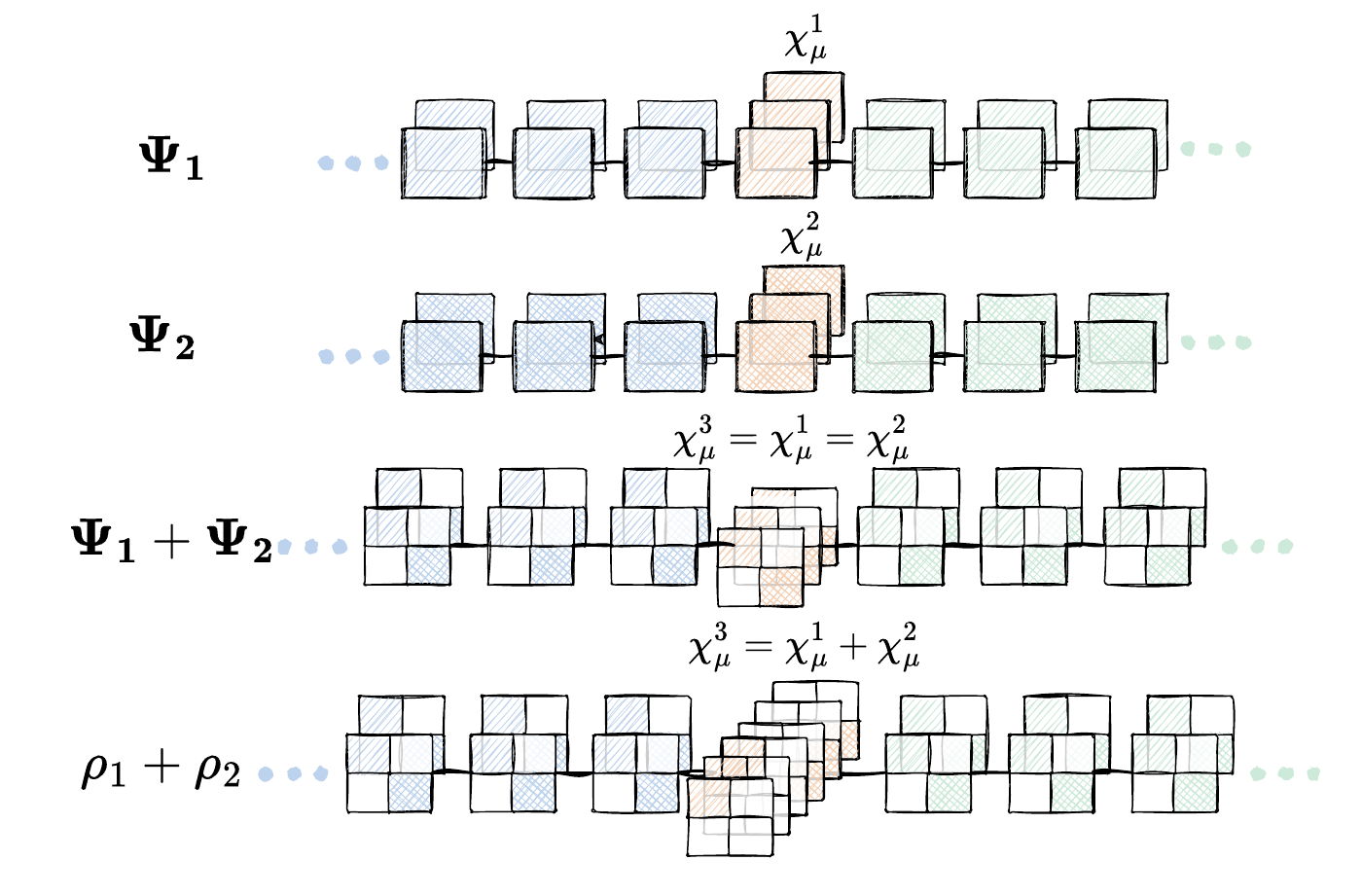}
    \caption{\textbf{Two different ways to sum two tensor networks.} Each matrix (square) appears in different versions depending on the value of the corresponding physical index. The blue and green squares represent the two different modes while the orange ones correspond to the purity tensor. The third (resp. fourth) row corresponds to vector (resp. matrix) addition. }
    \label{fig:2_sums}
\end{figure}

To perform vector addition, we use the standard addition rules for two MPS or tree tensor networks, treating the purification index $\mu$ as any other physical index.
We refer to e.g. \cite{Waintal2026} for a presentation of the vector addition algorithm. 
In both additions, the tensors $\Psi^{(\alpha)}(n)_1$ and $\Psi^{(\alpha)}(n)_2$
become 
\begin{equation}
\Psi^{(\alpha)}(n) = \begin{pmatrix} \Psi^{(\alpha)}(n)_1 & 0 \\ 0 & \Psi^{(\alpha)}(n)_2 \end{pmatrix}.
\end{equation}
(More precisely, one should write this equation with the
$M^{(\alpha,i)}(\sigma_i)$ tensors). The difference between the two additions lies in the way one treats
the tensors $\Psi^{(0)}(\mu)_1$ and $\Psi^{(0)}(\mu)_2$. For vector addition, the size $\chi_{\mu}$ remains unchanged, and this tensor 
becomes 
\begin{equation}
\Psi^{(0)}(\mu) = \begin{pmatrix} \Psi^{(0)}(\mu)_1 & 0 \\ 0 & \Psi^{(0)}(\mu)_2 \end{pmatrix}.
\end{equation}
In contrast, to perform matrix addition, the size $\chi_\mu$ becomes 
$\chi_{\mu1}+\chi_{\mu2}$ and we have
\begin{equation}
\Psi^{(0)}(\mu) = 
\begin{cases}
\begin{pmatrix} 
\Psi^{(0)}(\mu)_1 & 0 \\ 0 & 0 \end{pmatrix} 
\text{ if $1\le\mu \le \chi_{\mu1}$} \\
\begin{pmatrix} 0 & 0 \\ 0 & \Psi^{(0)}(\mu)_2 \end{pmatrix} 
\text{ if $\chi_{\mu1}<\mu \le \chi_{\mu1}+\chi_{\mu2}$}
\end{cases}
\end{equation}
Fig. \ref{fig:2_sums} shows a schematic of the two different kinds of additions in the case of 2 bosonic modes. Note that in both cases, a compression is performed after the addition.

\subsection{Lindblad integration as a Kraus map}\label{subsec:lucao}
We are now ready to turn to the Lindblad equation itself. An important remark is that we will not treat the Lindblad equation as an ordinary differential equation. Although, strictly speaking, it is one (if one treats the matrix $\rho$ as a large vector), this would jeopardize its additional structure and prevent us from taking advantage of the purification. Instead, we will write the integration scheme in terms of Kraus maps \cite{Nielsen2011}, following the schemes described in \cite{Cao2025}. 

\subsubsection{General Kraus operators algebra}
A Kraus map $\mathcal{A}$ is defined by a set of matrices $A_1$...$A_I$ (called Kraus operators) such that,
\begin{equation}
\mathcal{A}(\rho) = \sum_{i=1}^{I} A_i \rho A_i^\dagger.
\end{equation}
Such an operator can readily be applied to our representation. Indeed, in terms of our purification, each of the $I$ terms simply corresponds to a matrix vector multiplication $\boldsymbol{\Psi} \rightarrow A_i \boldsymbol{\Psi}$ (this is a MPO $\cdot$ MPS multiplication); the sum corresponds to the \emph{matrix} addition defined in the preceding subsection. Hence, given the $A_i$, we know how to compute $\mathcal{A}(\rho)$. If a matrix $A_i$ is itself a sum of matrices, say $A_1 = B+C$ then to apply $\mathcal{A}(\rho) = A_1 \rho A_1^\dagger$, one needs to compute 
$B \boldsymbol{\Psi}$ and $C \boldsymbol{\Psi}$ and then perform the \emph{vector} addition between these two tensor networks.

Kraus maps can be added 
\begin{equation}
[\mathcal{A}+\mathcal{B}](\rho) = \mathcal{A}(\rho) 
+ \mathcal{B}(\rho)
\end{equation}
(this is just matrix addition for $\boldsymbol{\Psi}$) and multiplied through the composition
\begin{equation}
[\mathcal{A}\mathcal{B}](\rho) = \mathcal{A}(\mathcal{B}(\rho))
\end{equation}
(this is just applying one, then the other). 
Since we know how to deal with them in our representation, our goal is therefore to find a Kraus map $\mathcal{K}$ such that,
\begin{equation}
\rho(t+h_t) \approx \mathcal{K}[ \rho (t)]
\end{equation}

\subsubsection{First and second order schemes}
Fortunately, this problem has already been solved in the literature.
The first step is to absorb part of the Lindblad operators as an effective (non-Hermitian) Hamiltonian. We rewrite Eq.\eqref{eq:lindblad} as,
\begin{equation}
    \frac{d\h{\rho}}{dt} = -i\left[ \h{H}_{\mathrm{eff}}\h{\rho} - \h{\rho}\h{H}_{\mathrm{eff}}^\dagger \right] + \mathcal{L}(\h{\rho})
\end{equation}
with 
\begin{equation}
\h{H}_{\mathrm{eff}} \equiv \h{H} -\frac{i}{2} \sum\limits_{k=1}^l \h{L}_k^\dagger \h{L}_k
\end{equation}
and the jump operator,
\begin{equation}
\mathcal{L}(\h{\rho}) \equiv \sum\limits_{k=1}^l \h{L}_k \h{\rho}\h{L}_k^\dagger.
\end{equation}
We further define the Kraus map $\mathcal{U}$ that consists of a single
matrix $U_1$,
\begin{equation}
U_1 = 1 - i h_t \h{H}_\text{eff}
\end{equation}
as well as the Kraus map $\mathcal{U}'$ that consists of a single
matrix $U'_1$,
\begin{equation}
U'_1 = 1 - i h_t \h{H}_\text{eff} - \tfrac{h_t^2}{2} \h{H}_\text{eff}^2
\end{equation}
The simplest integration scheme is the first order Kraus map \cite{Haroche2006},
valid up to second order corrections:
\begin{equation}
\mathcal{K} = \mathcal{U}  + h_t \mathcal{L}.
\end{equation}
\cite{Cao2025} introduces integration schemes valid to any order. In this work, we mostly use the second order scheme, valid up to $O(h_t^3)$ corrections, that reads,
\begin{equation}
\mathcal{K} = \mathcal{U}'  + \tfrac{h_t}{2} [\mathcal{U}\mathcal{L}
+\mathcal{L}\mathcal{U}] + \tfrac{h_t^2}{2} \mathcal{L}^2.
\end{equation}
These schemes do not preserve the trace of the density matrix exactly.
They can be made trace preserving by adding a renormalization step $\rho \rightarrow\rho/\text{Tr}(\rho)$ after each time step.

\section{Application: Z gate on a large cat qubit}\label{sec:cat}

The first application of interest for our method is the simulation of the evolution of a cat qubit \cite{Mirrahimi2014,Guillaud2023} undergoing a Z gate. A cat qubit consists of two superconducting microwave cavities (quantum harmonic modes) coupled via a nonlinear element called the ATS \cite{Lescanne2020}, composed of two Josephson junctions shunted by an inductance. Under appropriate experimental conditions, the ATS mediates a two-to-one photon conversion process, in which pairs of photons in one cavity (the memory, or cat cavity) are converted into single photons in the other cavity (the buffer) at a rate $g_2$. The buffer is engineered to be strongly dissipative and is driven resonantly with amplitude $\varepsilon_d$. After adiabatic elimination of the ATS, and including losses in both the memory and buffer modes, the system Hamiltonian in the rotating frame of both modes can be written as:

\begin{equation} \label{eq:cb}
    \left\{\begin{array}{l}
    \h{H}_\text{stab} = g_2 \h{a}^{\dagger 2} \h{b}+g_2^* \h{a}^2 \h{b}^{\dagger}+\epsilon_d\left(\h{b}+\h{b}^{\dagger}\right)  \\
    \h{L}_b=\sqrt{\kappa_b} \h{b} \\
    \h{L}_a=\sqrt{\kappa_a} \h{a}
    \end{array}\right.
\end{equation}

Where $\h{a}$ (resp. $\h{b}$) is the annihilation operator on the memory cavity (resp. the buffer cavity). To simulate a Z gate, we must add a drive on the cavity for a duration $T_Z$. In the rotating frame, the drive of amplitude $\varepsilon_Z$ has the form,
\begin{equation}\label{eq:Hz}
    \h{H}_z = \varepsilon_Z (\h{a} + \h{a}^\dagger)
\end{equation} 

\begin{figure*}[ht!]
    \centering
    \includegraphics[width=\textwidth]{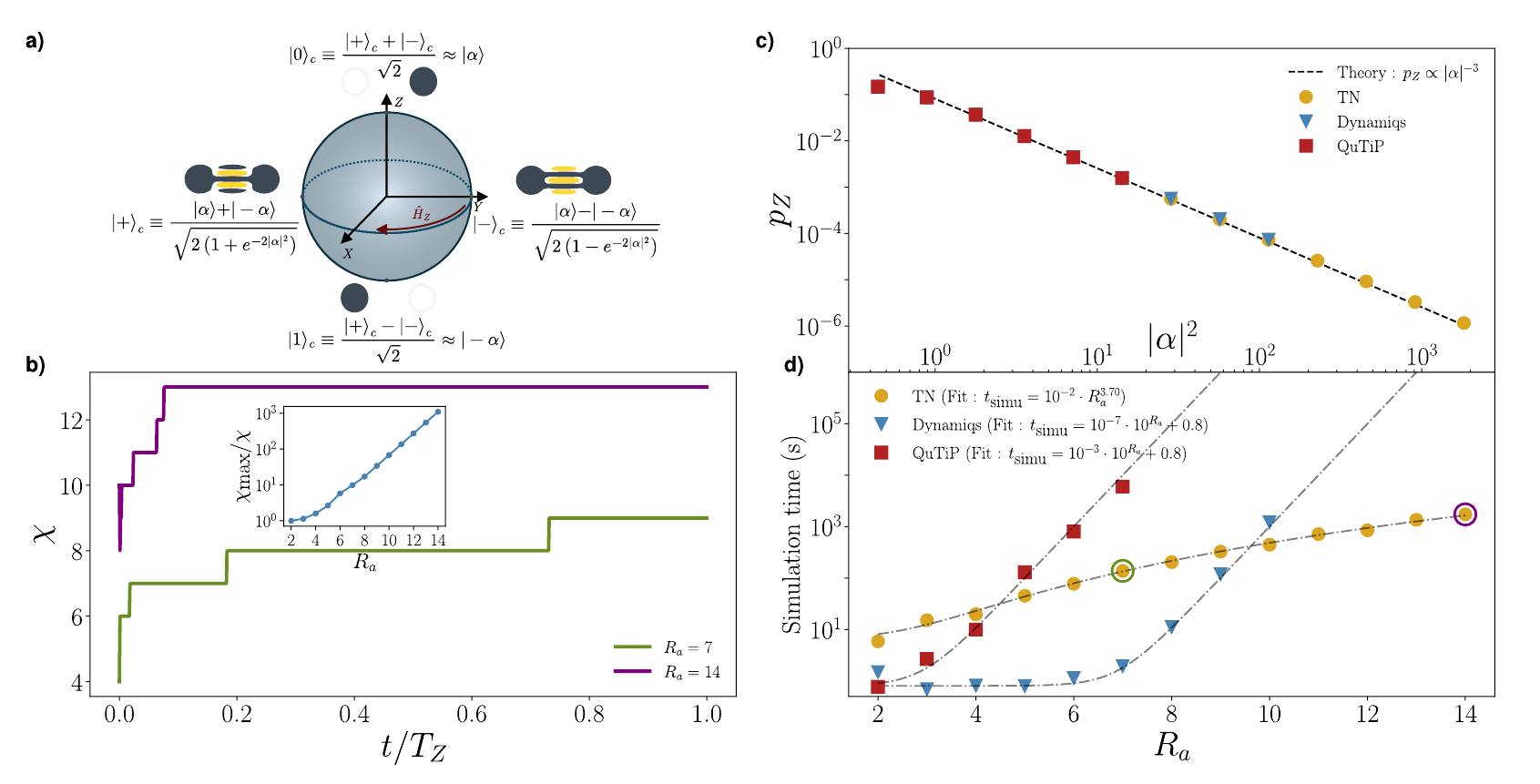}
    \caption{\textbf{Z-gate simulation a)} Bloch sphere of a cat qubit. \textbf{b)} Evolution of the quantics bond dimension of the TN during the simulation, for two values of $R_a$. The inset shows the compression factor defined as $\chi_{\text{max}} /\chi \equiv 2^{R_a} /\chi $. \textbf{c)} Error rate of the gate, computed with both our ansatz and a full-rank simulator, as a function of the size of the cat qubit, with $|\alpha|^2$ ranging from $0.5$ to $2000$. \textbf{d)} Evolution of the simulation time as a function of $|\alpha|^2 = 2^{R_a}/3$. Dotted lines are fitted functions demonstrating the scaling. We observe a polynomial scaling of the simulation time with the truncation size for the naïve scheme, while our ansatz scales as a logarithm of the truncation size. The two circles correspond to the two points studied in panel b.
    }
    \label{fig:Z_gate}
    
\end{figure*}

\subsection{Elements of cat qubit theory}
The Lindblad equation described by the first two lines of Eq.\eqref{eq:cb} stabilizes the two computational states: 
\begin{equation}
    \ket{\psi} = \ket{\pm \alpha}_a \bigotimes \ket{0}_b \text{ with } \alpha = -\sqrt{\frac{\varepsilon_d}{g_2}}
\end{equation}
The experimental parameters are often chosen such that the system is in the adiabatic regime, with 
\begin{equation}
    8\alpha g_2 \ll \kappa_b
\end{equation}
so that the buffer cavity remains mostly empty at all times.

The main appeal of a cat qubit is that it is inherently protected against bit-flip errors. Indeed, when one considers only a single photon loss operator 
($\h{L}_a=\sqrt{\kappa_a} \h{a}$), the probability of bit-flip 
$p_{\text{b-f}}$ is exponentially suppressed upon increasing the size of the cat $|\alpha|^2$ \cite{Lescanne2020,Reglade2024, Dubovitskii2025}. This however comes at the cost of a linear increase of the probability of the phase-flip $p_{\text{p-f}}$. 
In short the theory predicts,
\begin{equation} \label{eq:errors}
    \left\{\begin{array}{l}
    p_{\text{b-f}} \varpropto e^{-c |\alpha|^2}  \\
    p_{\text{p-f}} \varpropto |\alpha|^2  \\
    \end{array}\right.
\end{equation}
A Z gate is a rotation of the cat qubit along its Z-axis. This operation is done by adding a drive on the memory (see Eq.\eqref{eq:Hz})
\begin{equation}
    \h{H} = \h{H}_\text{stab} + \h{H}_Z
\end{equation} 
This has the effect of rotating the qubit along its Z-axis, hence a Z gate of arbitrary angle $\theta$ can be realized, with ${\theta = 4\alpha T_Z \varepsilon_Z }$. See Fig. \ref{fig:Z_gate}a for a schematic illustration. During this gate, two types of processes can lead to phase-flip errors (also known as Z errors):
\begin{itemize}
    \item Photon loss of the qubit during the gate:
    \begin{equation}
        p_{\text{ph}} \equiv |\alpha|^2\kappa_aT_Z
    \end{equation}
    \item The drive on the memory induces a displacement of the buffer mode that depends on the state of the memory. Since the buffer cavity is very lossy, this information is then lost to the environment, which induces phase errors \cite{Gautier2023}. The expression for this error is given by:
    \begin{equation}
        p_{\text{buf}} \equiv \frac{\theta^2}{16 |\alpha|^4 T_Z} \frac{\kappa_b}{4(g_2)^{2}}
    \end{equation}
\end{itemize}
So in the end we expect the total Z gate error to be equal to:
\begin{equation}
    p_Z \equiv p_\text{p-f} = p_{\text{ph}} + p_{\text{buf}}
\end{equation}

\subsection{Numerical experiment}
We consider a system with the following parameters: $g_2 = 1$, $\kappa_b = 10g_2$, $\varepsilon_Z = \frac{1}{10}\frac{4(g_2{})^2}{\kappa_b}$, $\theta = \pi$ and we neglect photon loss, $\kappa_a = 0$. We choose here to place ourselves in a regime where the errors caused by single photon loss are negligible in order to concentrate on the contribution from the buffer. For the chosen parameters, we expect
\begin{equation}
\label{eq:pZ}
    p_Z = \frac{\pi}{40|\alpha|^3}
\end{equation}
In this benchmark, we compare our simulations (with tensor networks) with two full-vector simulator packages, QuTiP \cite{Lambert2026} which runs on CPUs and Dynamiqs \cite{Guilmin2025} which is optimized for GPUs.
To show the scaling of the method, we vary the Hilbert space size of the memory $N_a = 2^{R_a}$ while varying the size of the cat qubit $|\alpha |^2 \equiv N_a/3$. We take a fixed value of $R_b = 4$ ($N_b = 2^{R_b} = 16$) for the buffer. Fig. \ref{fig:Z_gate} presents the results of our simulations. In Fig. \ref{fig:Z_gate}c, we observe that we accurately obtain the error rate Eq.\eqref{eq:pZ} predicted by the theory. More importantly, in Fig. \ref{fig:Z_gate}d, we note that the simulation time behaves differently from the naïve method. Indeed, in the full-vector simulations we get a polynomial scaling in the system size (exponential in the number of tensors), whereas with our ansatz, the simulation time only scales logarithmically with the system size (polynomial in the number of tensors). With this, we obtain a significant gain of simulation time for large systems. For the largest Hilbert space size considered, our simulations ran in under an hour with a memory footprint of 100 kB compared to an expected simulation time of half a year (resp. 3 millennia) and a memory footprint of 512 GB (resp. 512 GB) for the Dynamiqs (resp. QuTiP) software. Also, note that the Dynamiqs simulations were carried out on a GPU, while the TN simulations were done using only a single CPU. The benchmark is thus heavily favored towards Dynamiqs. In a fairer benchmark on the same hardware, the crossover between the two methods occurs instead at $|\alpha|^2 \approx 3$. This major improvement can be explained with Fig. \ref{fig:Z_gate}b that illustrates the evolution of the bond dimension through time for various Hilbert-space sizes and for a precision of $10^{-8}$. We observe that it plateaus to a constant value after a certain point of the evolution. This demonstrates that the state of the system remains very compressible during the whole evolution, and therefore the simulation time remains bounded and does not blow up with the truncation size.

\section{Application: Transmon Ionization }\label{sec:transmon}

\subsection{Strongly driven transmon}
We now turn our attention toward a second problem, the so-called transmon ionization, which is typically hard to simulate as it requires a large Hilbert space to capture the ionization process. Over the years, the transmon has emerged as one of the most promising platforms to use as a qubit \cite{Koch2007}. It consists of a Josephson Junction in parallel with a capacitance. The transmon Hamiltonian takes the form (in the absence of charge noise):
\begin{equation}
    \label{eq:transmon}
    \h{H}_\text{t} = 4E_C\h{n}_\text{t}^2 - E_\text{J}\cos{(\h{\phi}_\text{t})}
\end{equation}
In the charge basis (beware, this is \emph{not} the Fock basis), we have 
\begin{align}
\h{n}_\text{t} &= \sum_{n=-\infty}^\infty n \ketbra{n}{n} \\
\cos{(\h{\phi}_\text{t})} &= \frac{1}{2} \sum_{n=-\infty}^\infty \ketbra{n}{n+1} + \ketbra{n+1}{n}
\end{align}
In its usual operation mode, only the two lowest eigenstates of this Hamiltonian are used as logical states, the $\ket{0}$ and $\ket{1}$.
The transmon readout is usually performed by coupling the transmon capacitively to a resonator (hereafter, the "cavity" $c$), which is initially empty \cite{Blais2021}. Then, applying a drive of amplitude $\varepsilon_d$ and frequency $\omega_d$ on the cavity will make it evolve to two different states depending on the state of the transmon. 
Calling $\h{a}$ the cavity annihilation operator, the total Hamiltonian of the system is given by:
\begin{align}
\begin{split}
\label{eq:transmon_cavity}
    \h{H}(t) &=  4E_C\h{n}_\text{t}^2 - E_\text{J} \cos{(\h{\phi}_\text{t})} + \omega_r\h{a}^\dagger\h{a} \\
     &- ig\h{n}_\text{t}(\h{a} - \h{a}^\dagger) -i\varepsilon_d \sin{(\omega_d t)}(\h{a} - \h{a}^\dagger)
\end{split}
\end{align}
to which we add a Lindblad operator $\h{L} = \sqrt{\kappa} \h{a}$ to account for the photon losses in the cavity at a rate $\kappa$.
The readout speed of the transmon is directly linked to the strength of the drive applied to the cavity. Indeed, for moderate drive amplitudes, increasing the drive amplitude yields a faster readout. However, if the drive is too strong, it can generate undesired effects that affect the fidelity of the measurement as well as its QNDness \cite{Boissonneault2009}. One of those effects is called the transmon ionization \cite{Shillito2022}. This effect occurs when states outside the potential well of the transmon get populated during the evolution because of spurious transitions. Transmon ionization can be predicted at a qualitative level by performing a branch analysis of \cite{Shillito2022} that we reproduce in Appendix \ref{sec:branch} for completeness. 

\begin{figure*}[ht!]
    \centering
    \includegraphics[width=\textwidth]{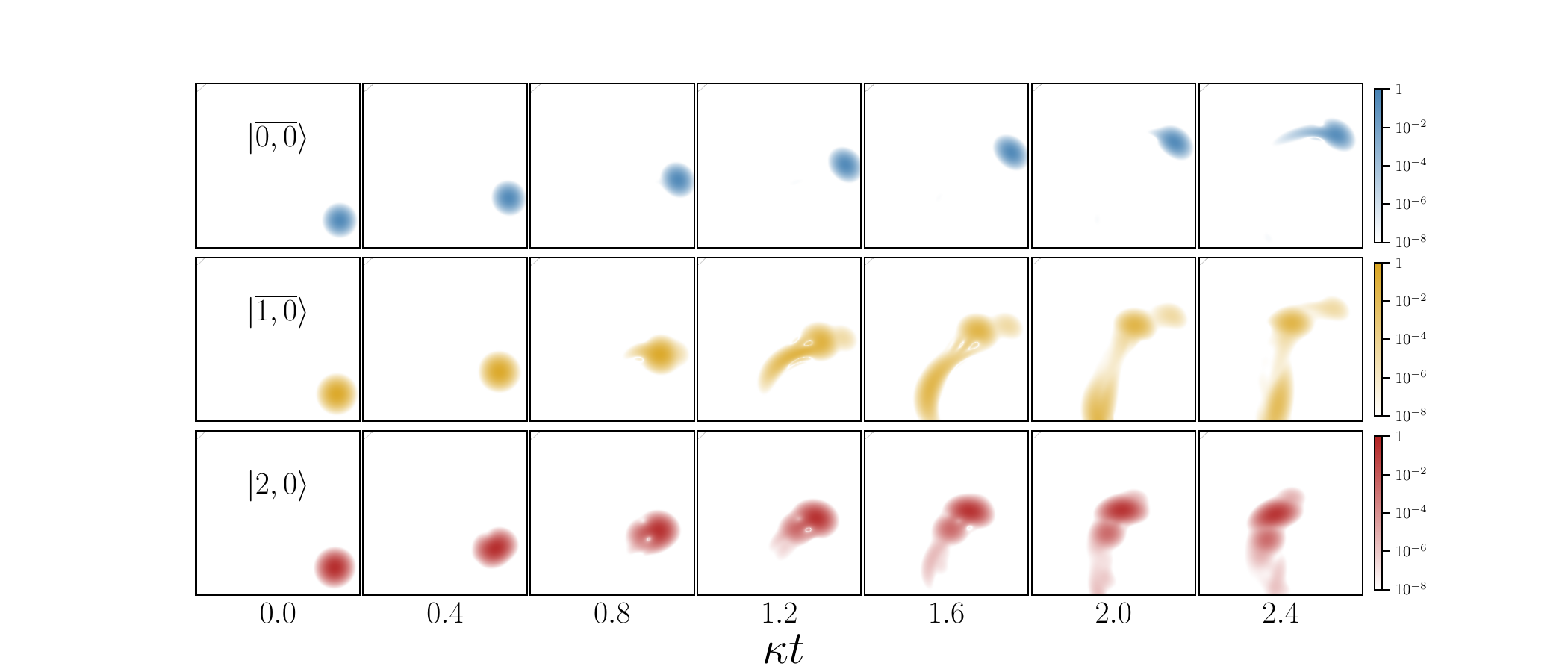}
    \caption{\textbf{Wigner function of the cavity at different times for the Hamiltonian of Eq.\eqref{eq:transmon_cavity}.} 
    Each colormap shows the Wigner function in the $(\Re(\alpha),\Im(\alpha))$ plane in logarithmic scale for $\varepsilon_d/2\pi = 280\text{ MHz}$ and a given time. 
 The panels correspond to three initial states $\ket{\overline{0,0}}$ (blue, top),
    $\ket{\overline{1,0}}$ (orange, middle) and
    $\ket{\overline{2,0}}$ (red, bottom). 
    We use  $\Re(\alpha) \in [-13,2]$ and $\Im(\alpha)\in [0,15]$.}
    \label{fig:wigners}
\end{figure*}

In this study, we use the parameters $E_C/h = 280 \text{ MHz}$, $E_J/E_C = 50$, $\omega_r/2\pi = 7.5 \text{ GHz}$, $g/2\pi = 250 \text{ MHz}$ and $ \kappa/2\pi = 20 \text{ MHz}$. This gives a transmon frequency
$\omega_t/2\pi \approx \sqrt{8E_CE_J} - E_C \approx 5.320$ GHz. For those parameters we expect to see a different behavior depending on the initial state of the transmon, see the branch analysis in Fig. \ref{fig:branch_transmon} in Appendix \ref{sec:branch} for more details. Large truncations are required for both the transmon, because of interactions between qubit states and high-energy states above the potential well, and the cavity, because large photon numbers are reached when the ionization occurs. In the original paper that has motivated this section \cite{Shillito2022}, the authors used a total Hilbert space of size up to $2^{15}$, so that the density matrix had around one billion elements. To be able to perform the corresponding simulations, the authors had to use Google's Tensor Processor Units (TPUs) that provided around one PFLOP of computation power. With our ansatz, we were able to reproduce the results of their simulation using a machine with a single CPU, running for a few hours. The device that we used had a peak computation power of about 10 GFLOP which therefore corresponds to a factor $10^5$ reduction of the required computational resources.

\begin{figure*}[ht!]
    \centering
    \includegraphics[width=\textwidth]{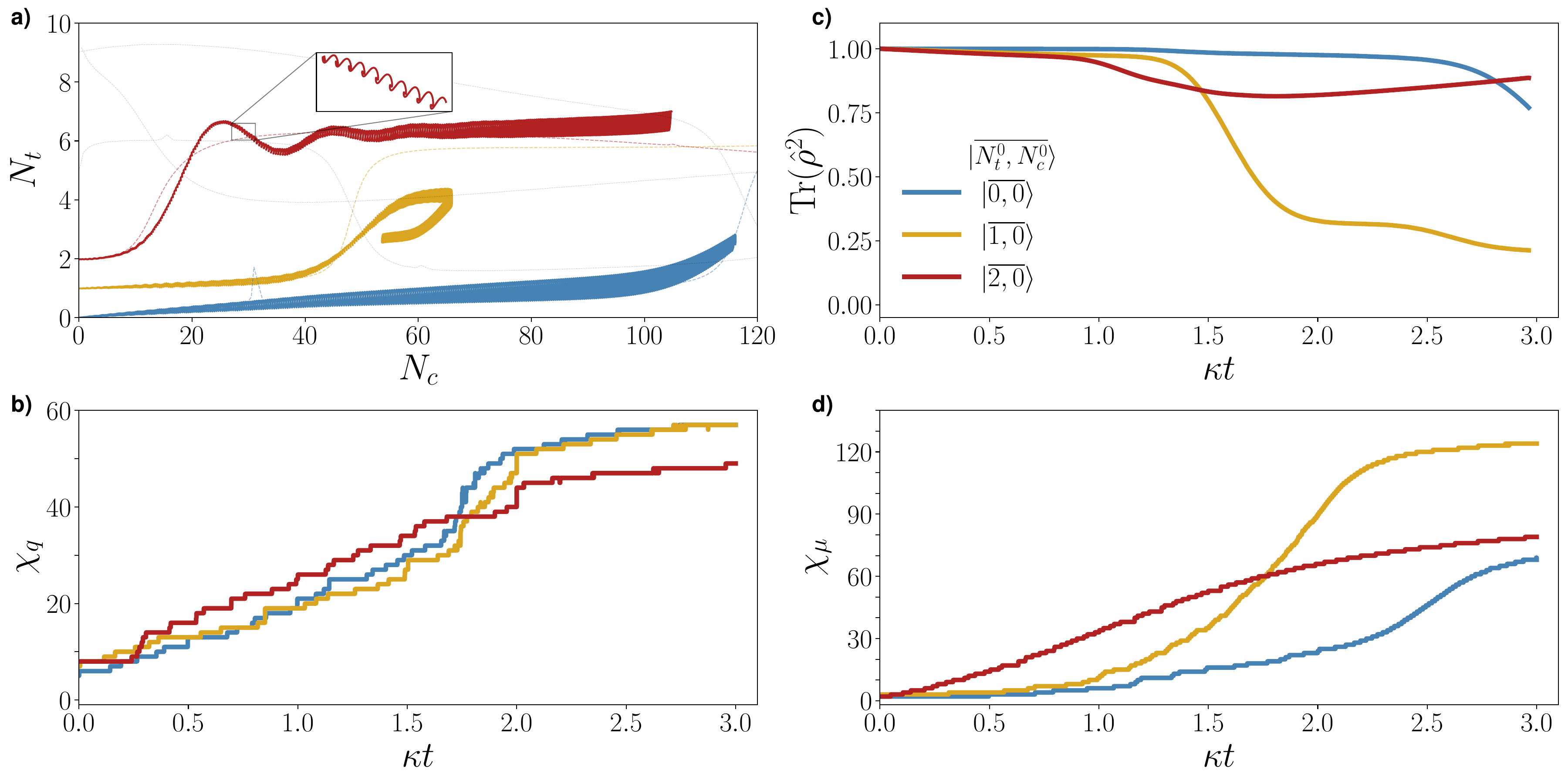}
    \caption{\textbf{Simulation of the dynamics of the Hamiltonian Eq.\eqref{eq:transmon_cavity} with} $\varepsilon_d/2\pi = 280\text{ MHz}$ 
    for various initial states. 
    \textbf{a)} Parametric plot of the population of the transmon against that of the cavity during the evolution. The inset shows the rapid oscillations during the evolution. The dotted lines correspond to selected branches of the undriven Hamiltonian. The trajectories for the first two dressed states (blue and orange) match the ones extracted from \cite{Shillito2022}, while the third (red) is novel.
    \textbf{b)} Evolution of the quantics bond dimension $\chi_q$. \textbf{c)} Evolution of the purity of the system. The ionization process is associated with a loss of purity, with varying amplitude depending on the initial state.
    \textbf{d)} Evolution of the purity bond dimension $\chi_\mu$. The ionization is associated with an increase of $\chi_\mu$ when it occurs. }
    \label{fig:transmon}
\end{figure*}
\subsection{Numerical experiment}
For this simulation, the anharmonicity of the transmon makes the Fock basis less natural than in the cat qubit case, which prompted us to use the eigenbasis of the transmon instead of the Fock basis. 
The different steps to obtain the MPO of the transmon Hamiltonian are as follows. First, we diagonalize the transmon Hamiltonian Eq.\eqref{eq:transmon} alone using a large cutoff for the charge $|n|\le 500$. This gives orthonormal eigenstates $\Psi_\alpha(n)$ and the corresponding eigenenergies $\varepsilon_\alpha$, of which we keep only the first 
$2^{R_t}$ states. Second, we write $\alpha$ in quantics form and use
TCI to obtain the diagonal MPO of $\hat H_t$ in this basis. Third, for the coupling term, we rewrite $\h{n}_t$ in the $\alpha$ basis,
\begin{equation}
[\hat n_t]_{\alpha\beta} \equiv \sum_{n}\Psi^*_{\alpha}(n) n \Psi_{\beta}(n)
\end{equation}
Fourth, we give the result of this expression to TCI to obtain the corresponding low-rank Quantics MPO representation. We also build the number operator $\h{b}^\dagger \h{b}$ in this basis where $\hat{b}$ is the annihilation operator for the transmon by using the techniques described in \ref{sec:mpo_build}. We find numerically that all those operators admit a low rank representation in Quantics formalism in this basis with a typical quantics bond dimension of less than 10.

The transmon is a slightly anharmonic oscillator. Ignoring this anharmonicity and the transmon-cavity coupling, the occupation numbers of the transmon ($N_t$) and cavity ($N_c$),
\begin{eqnarray}
N_t &\equiv& \Tr(\h{b}^\dagger \h{b} \h{\rho}) \\
N_c &\equiv& \Tr(\h{a}^\dagger \h{a} \h{\rho})
\end{eqnarray}
would be good quantum numbers that could be used to label each eigenstate
$\ket{i_t,i_c}$ with $i_t=N_t$ and $i_c=N_c$. Now, in the presence of the anharmonicity and the coupling,
we can still use these two numbers to label the eigenstates of the total
Hamiltonian without drive $\h{H}(t=0)$ (see Appendix \ref{sec:branch} for more details on how this labeling is performed in practice). We call
these dressed states $\ket{\overline{i_t,i_c}}$ and 
$\omega_{\overline{i_t,i_c}}$ the associated energy.
To place ourselves in the regime used experimentally for a readout, the drive frequency is set at a value given by the average of the cavity frequency over the transmon states,
\begin{equation}
    \omega_d = \tilde{\omega}_r\equiv \frac{1}{2}
    \sum_{i_t=0}^1 \omega_{\overline{i_t,1}} - \omega_{\overline{i_t,0}}
\end{equation} 
For our set of parameters, this yields $\omega_d = 7.522$ GHz.
We simulate the readout process, where the system is initially either in the logical state $\ket{\overline{0,0}}$ or $\ket{\overline{1,0}}$. We also look at the behavior of the system starting in state $\ket{\overline{2,0}}$
(leakage). In order to accurately capture the high-energy phenomena, we use a Hilbert space size of $2^{10}$ for the cavity and $2^5$ for the transmon. Finally, we use a relative tolerance of $10^{-8}$ for all operations of compression on the tensor network. The drive amplitude is set to 
$\varepsilon_d/2\pi = 280 \text{ MHz}$.

Fig. \ref{fig:wigners} shows the evolution of the Wigner function \cite{Hillery1984} of the cavity during the readout process. 
In a standard readout, the spot should go up and then eventually separate and turn right or left depending on the state of the transmon.
When the system starts in its ground state, the cavity gets populated during the evolution but remains close to a coherent state (upper panels, the spot remains approximately circular). However, when the transmon is initially in one of the excited states, the Wigner function gets "bananified" (middle and lower panels). These new features are an indication of a problem in the readout with highly excited states of the transmon getting populated. 

Fig. \ref{fig:transmon}a shows a parametric plot of the population of the transmon ($N_t$) against that of the cavity ($N_c$) during the evolution, for various initial states. 
We observe that the dynamics follows closely the one predicted by the branch analysis (thin gray lines in the background, see Appendix \ref{sec:branch}), with additional oscillations caused by the driving frequency being different from the resonant frequency of the system (see the zoom in the inset). 
The blue curve (initial state in $\ket{\overline{0,0}}$) shows a slow 
steady increase of the transmon population as the cavity population increases. However, the orange curve (initial state in $\ket{\overline{1,0}}$) shows the beginning of a transmon ionization process when $N_c\approx 50$
with a rather rapid jump of the transmon population. This is associated with a sudden loss of purity, as shown in Fig. \ref{fig:transmon}c around 
$\kappa t\approx 1.7$.

From a numerical perspective, the initial state is a superposition of a few eigenstates of the Hamiltonian, so it is pure and of low rank. In Fig. \ref{fig:transmon}b, we show the evolution of the quantics bond dimension during the evolution, we find that it remains under $\chi_q <60$ at all times, much smaller than its maximum possible value of $2^{10}$; also it reaches a plateau value after which it stops increasing. Similarly, in Fig. \ref{fig:transmon}d the purity bond dimension $\chi_\mu$ increases when the ionization process occurs (Fig. \ref{fig:transmon}c), as it is associated with an increase of the mixedness of the system. However, $\chi_\mu\approx 120$ remains much smaller than its maximum possible value of $2^{15} \approx 32, 000$. This demonstrates that even though this system undergoes a complex evolution that leads to a highly mixed state, its density matrix can still be efficiently compressed using our ansatz; this is relevant because this system essentially breaks all the assumptions needed to get a large compression ratio. Even in this situation, we could represent the state in a low-rank format that lets us evolve it efficiently on a single CPU core, i.e.  orders of magnitude less computing power than the TPUs used originally.

\section{Conclusion}\label{sec:conclusion}
We presented a method to simulate Lindblad dynamics of bosonic quantum systems by combining Tensor Train representations with quantics discretization and Tensor Cross Interpolation. The compression of the density matrix exploits three complementary axes: the purity of the state, the low entanglement between subsystems, and the QTT representation of each local Hilbert space. Together, these provided a drastic reduction in the memory footprint of the density matrix, as illustrated in Fig. \ref{fig:compression_cat}.

 For closed systems, we found that TDVP with a fourth-order Magnus expansion gives the best accuracy while keeping bond dimensions low, outperforming standard methods when the Hamiltonian is time-dependent. For open systems, we described numerical schemes adapted to this representation, based on Kraus maps, and showed how they integrate naturally into the tensor network formalism.

We applied this method to two problems: a Z gate on large cat qubits and transmon ionization under a strong cavity drive. In both cases, the bond dimension remains well below its theoretical maximum throughout the evolution, which explains the favorable scaling with system size observed in Fig. \ref{fig:Z_gate}d. Notably, the transmon ionization simulation, which originally required Google TPUs, was reproduced on a single CPU core, reducing the required computational resources by five orders of magnitude.

In a field where experimental devices are getting larger and more complex, this method provides a direct path to simulate accurately the behavior of such devices while mitigating the curse of dimensionality. Further work should be dedicated to applying this ansatz to larger and more complex systems to evaluate its capacity to accurately represent such regimes.

\section{Acknowledgments}\label{sec:acknowledgments}
We thank Christoph Groth, who helped review some of our code. We also thank Jheng-Wei Li and Nicolas Jolly for helpful discussions. X.W. acknowledges fundings from Plan France 2030 ANR-22-PETQ-0007 “EPIQ”, the PEPR “EQUBITFLY”, the ANR “DADI”, the ANR TKONDO and the CEA-FZJ French-German project AIDAS. A.M. would like to express his thanks to his colleagues at Alice \& Bob who accompanied him throughout this project, especially Emilio Rui, Patrick Winkel, Nora Reinic and Nicolas Didier.

\appendix
\section{Branch analysis}\label{sec:branch}

\begin{figure*}[ht!]
    \centering
    \includegraphics[width=.9\textwidth]{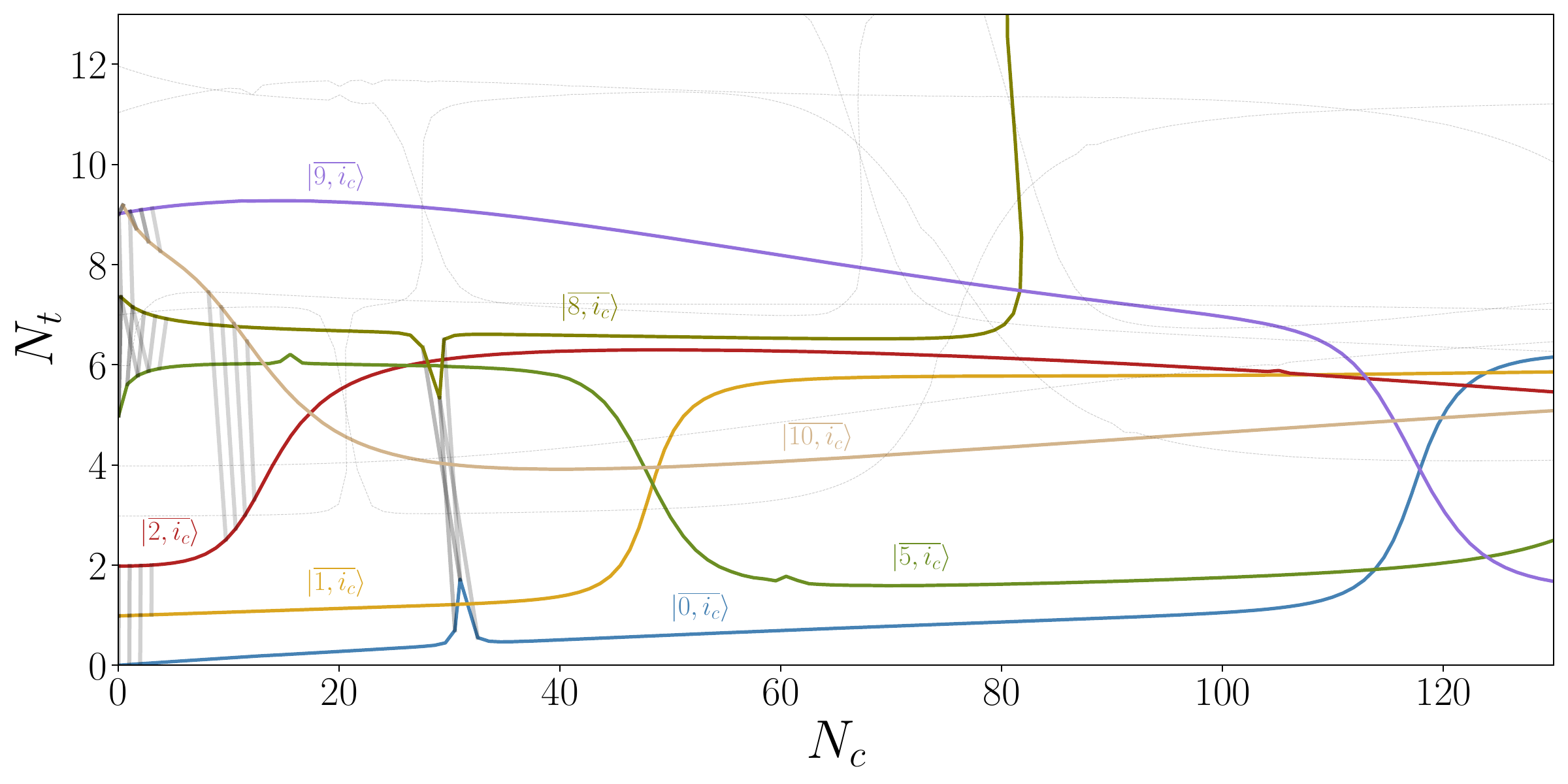}
    \caption{\textbf{Parametric plot of the eigenstates of the Hamiltonian in the $(N_t,N_c)$ plane.} The different colors correspond
    to $\ket{\overline{0,i_c}}$ (blue), $\ket{\overline{1,i_c}}$ (orange),
    $\ket{\overline{2,i_c}}$ (red) and so on.
     The nearly vertical dark gray lines indicate that the identification of the branch is ill-defined and that the state has a significant overlap with the other branch connected by the gray line.}
    \label{fig:branch_transmon}
\end{figure*}

In this section, we summarize the branch analysis method used in \cite{Shillito2022, Dumas2024} to study the Hamiltonian of the transmon qubit in the absence of drive. The starting point is the set of eigenenergies $\epsilon_\alpha$ and corresponding eigenstates $\Psi_\alpha$.
The goal is, given the Hamiltonian (truncated at a certain Hilbert space size) and its eigenstates $\{\ket{\lambda_j}\}_{j=0...N}$, to identify the dressed states, i.e. the eigenstates of the Hamiltonian $\ket{\overline{i_t,i_c}}$. To perform this identification, we start with the identification of the empty cavity states, $i_c=0$. A state $\ket{\Psi_\alpha}$ (with $N_c\approx 0$) is identified with the state $\ket{i_t,0}$ with which the overlap of the corresponding state is the highest, i.e. $\ket{\Psi_\alpha}= \ket{\overline{i_t,0}}$ with
\begin{equation}
i_t(\alpha) = \underset{i_t} {\text{argmax}}
 |\langle \Psi_\alpha \ket{i_t,0}|^2
\end{equation}
To continue, we build the branches iteratively,
by identifying $\ket{\overline{i_t,i_c+1}}$ with the (yet unlabeled) state that has the maximum overlap with $\h{a}^\dagger\ket{\overline{i_t,i_c}}$,
\begin{equation}
    \ket{\overline{i_t,i_c+1}} \equiv 
    \underset{\text{unlabeled\ }\ket{\Psi_\alpha}}{\text{argmax}} 
    |\langle \Psi_\alpha|\h{a}^\dagger\ket{\overline{i_t,i_c}}|^2
\end{equation}

Fig. \ref{fig:branch_transmon} shows a reproduction of the branch analysis from \cite{Shillito2022}. For each branch, we plot parametrically the cavity population, given by $N_c = \bra{\overline{i_t,i_c}}\h{a}^\dagger \h{a}\ket{\overline{i_t,i_c}}$, against the transmon population, given by $N_t = \bra{\overline{i_t,i_c}}\h{b}^\dagger \h{b}\ket{\overline{i_t,i_c}}$ for all eigenstates with the different colors indicating the labeling that has been done. In an ideal case, we should observe that the spectrum is made of perfectly horizontal lines. This is indeed the case for small cavity population. For $N_c\gg 1$, the behavior depends on the transmon population quite strongly. As one drives the cavity, which is going to increase $N_c$,
we expect different behavior depending on whether we start from 
$\ket{\overline{0,0}}$ or $\ket{\overline{1,0}}$.
In the first case, we expect that the population of the transmon should slowly build up as we populate the cavity (blue).
However, if we start in $\ket{\overline{1,0}}$ (orange) or $\ket{\overline{2,0}}$ (red), we expect the population of the transmon to quickly increase as there is a crossing with a highly populated branch. As shown in \cite{Shillito2022}, this shift of population is expected to come with a loss of purity.

\bibliographystyle{apsrev4-2}
\bibliography{biblio}
\end{document}